\documentclass[]{emulateapj}
\usepackage{amsmath}

\shorttitle{The nitrogen carrier in protoplanetary disks}
\shortauthors{Pontoppidan et al.}

\begin{document}


\title{The nitrogen carrier in inner protoplanetary disks}


\author{Klaus M. Pontoppidan}
\affil{Space Telescope Science Institute, 3700 San Martin Drive, Baltimore, MD 21218, USA}
\email{pontoppi@stsci.edu}

\author{Colette Salyk}
\affil{Vassar College Physics and Astronomy Department, 124 Raymond Avenue, Poughkeepsie, NY 12604, USA}

\author{Andrea Banzatti}
\affil{Lunar and Planetary Laboratory, The University of Arizona, Tucson, AZ, 85721, USA}

\author{Geoffrey A. Blake}
\affil{Division of Geological and Planetary Sciences, California Institute of Technology, MC 150-21, 1200 E California Blvd., Pasadena, CA 91125, USA}

\author{Catherine Walsh}
\affil{School of Physics and Astronomy, University of Leeds, Leeds, LS2 9JT, UK}

\author{John H. Lacy}
\affil{Department of Astronomy, The University of Texas at Austin, 2515 Speedway, Stop C1400, Austin, TX 78712, USA}
\email{}

\author{Matthew J. Richter}
\affil{Department of Physics, University of California Davis, 1 Shields Avenue, Davis, CA 95616, USA}
\email{}



\begin{abstract}
The dominant reservoirs of elemental nitrogen in protoplanetary disks have not yet been observationally identified. Likely candidates are HCN, NH$_3$ and N$_2$. The relative abundances of these carriers determine the composition of planetesimals as a function of disk radius due to strong differences in their volatility. A significant sequestration of nitrogen in carriers less volatile than N$_2$ is likely required to deliver even small amounts of nitrogen to the Earth and potentially habitable exo-planets. While HCN has been detected in small amounts in inner disks ($<10\,$au), so far only relatively insensitive upper limits on inner disk NH$_3$ have been obtained. We present new Gemini-TEXES high resolution spectroscopy of the 10.75 \,$\mu$m band of warm NH$_3$, and use 2-dimensional radiative transfer modeling to improve previous upper limits by an order of magnitude to $\rm [NH_3/H_{nuc}]<10^{-7}$ at 1\,au. These NH$_3$ abundances are significantly lower than those typical for ices in circumstellar envelopes ($[{\rm NH_3/H_{nuc}}]\sim 3\times 10^{-6}$). We also consistently retrieve the inner disk HCN gas abundances using archival Spitzer spectra, and derive upper limits on the HCN ice abundance in protostellar envelopes using archival ground-based 4.7\,$\mu$m spectroscopy ([HCN$_{\rm ice}$]/[H$_2$O$_{\rm ice}$]$<1.5-9$\%). We identify the NH$_3$/HCN ratio as an indicator of chemical evolution in the disk, and use this ratio to suggest that inner disk nitrogen is efficiently converted from NH$_3$ to N$_2$, significantly increasing the volatility of nitrogen in planet-forming regions.
\end{abstract}



\section{Introduction}

Understanding the pathways of volatiles from the interstellar medium to planets is the subject of intense debate, with the Earth and the Solar system at its center, but with a growing generalization to exoplanetary systems \citep{Moriarty14,Ciesla15}. It remains unclear how the Earth obtained its primary volatiles (carbon, nitrogen and hydrogen/water), and the answer may include multiple sources, including a combination of cometary \citep{Greenwood11, Hartogh11} and chondritic late impactors \citep{Morbidelli00, Tartese13}. More broadly, both in-situ accretion, and the later delivery, of volatile elements to exoplanets is likely intimately linked to the chemistry of protoplanetary disks. The bulk molecular carriers of a given element during the formation of planetesimals determine its volatility; that is, how much of the element can be found in a solid form as a function of temperature, and therefore how much can be incorporated into planetesimals at a given location in the disk. Once sequestered into planetesimals, the volatiles may be delivered to forming terrestrial planets, either as a local process, or through later delivery during dynamical scattering events \citep{Morbidelli00}. 

The molecular environment in the inner regions of protoplanetary disks (within a few au) is likely fundamentally different from that of the outer, icy regions (10s of AU or more). As there is recent evidence, e.g., from submillimeter dust imaging \citep{Zhang18}, that some planets exist at large radii of 10-100 AU, we denote the inner $\sim 10\,$AU as the ``inner planet-forming region'', to indicate that this is where the Solar system formed its planets, and where almost all known exoplanets likely formed. 

\subsection{Nitrogen in the solar system}
Delivery of volatiles to terrestrial planets is an inefficient process. Indeed, carbon, nitrogen and oxygen (CNO) are highly abundant in volatiles in molecular clouds, prior to, and during, the star-formation process, with 10-35\% of elemental nitrogen accounted for in observations of ices \citep{Oberg11, Pontoppidan14}. The main carriers of nitrogen in molecular clouds and circumstellar envelopes are N$_2$, NH$_3$, HCN and possibly a carbon-dominated refractory dust component. In ices a minor amount of what is likely the OCN$^-$ ion is also sometimes seen.  

Conversely, nitrogen is highly depleted in the bulk Earth relative to its cosmic abundance, likely by as much as 5--6 orders of magnitude, and by 1--2 orders of magnitude relative to the chondritic N/H$_2$O abundance \citep{Marty12}. While it is uncertain how much nitrogen might be sequestered in the deep mantle, this uncertainty is unlikely to increase the terrestrial nitrogen abundance by more than a factor of two \citep{Halliday13}. In solar system comets, elemental nitrogen is also depleted, albeit by a somewhat smaller value of roughly two orders of magnitude \citep{Mumma11}. Further, the relative depletion of nitrogen in the Earth is much higher than that of carbon and hydrogen (as carried by water). Together, relative abundances of nitrogen in various solar system bodies suggest that the bulk molecular carriers of nitrogen in the Earth-forming disk material were significantly more volatile than water. N$_2$, for instance, is extremely volatile and generally only available for inclusion into solids below $\sim$15\,K for pure ice, and $\sim$25\,K in the case of N$_2$ frozen on water \citep{Bisschop06, Fayolle16}, corresponding to distances beyond 20-50\,au. Therefore, nitrogen present in the Earth and other terrestrial planets is likely delivered in the form of a less volatile carrier. 

\subsection{Nitrogen in the inner planet-forming region}

The dust temperatures in the inner regions of protoplanetary disks are sufficiently high that no ices will exist at any depth. In this region, the dominant CNO carriers will either be in the gas-phase or in a refractory component, stable to temperatures of at least a few 100\,K. Consequently, infrared observations of gas-phase lines from the most abundant nitrogen-bearing molecules will directly constrain major carriers within a few au. The most likely bulk carriers of nitrogen anywhere are NI and N$_2$. However, these species have no electric dipole or quadropole transitions, nor magnetic dipole transitions, and the atomic ground state has no hyperfine splittings, such that they are unobservable in the gas-phase. Their abundance can be crudely estimated as the fraction of the elemental abundance not accounted for by observation of secondary species, such as HCN and NH$_3$ \citep{Pontoppidan14}. 

Warm NH$_3$ in disks has not yet been detected. \cite{Mandell12} reported upper limits on warm NH$_3$ in inner disks of $[{\rm NH_3/H_2O}]<0.16-0.2$, based on high-resolution VLT-CRIRES spectra of transitions from the $\nu_1$ band around 3\,$\mu$m. Assuming an absolute water abundance of $\rm [H_2O/H_{\rm nuc}] = 5\times 10^{-5}$, this corresponds to a relatively weak upper limit on the absolute NH$_3$ abundance of $[{\rm NH_3/H}]<10^{-5}$ (where $n(\mathrm{H_{nuc}})=n(\mathrm{H})+2n(\mathrm{H}_2)$). For instance, this limit would not have detected NH$_3$ in concentrations similar to those in interstellar ices. NH$_3$ has some of its strongest warm bands in 8--13\,$\mu$m regions, of which some were observable with the $R=600$ high-resolution mode of Spitzer-IRS. Using these bands, \cite{Salyk11} reported stronger, but uncertain, upper limits of $[{\rm NH_3/H_2O}]<0.01$, based on a simple slab model and assuming an NH$_3$ gas temperature of 400\,K. If accurate, these limits suggest that NH$_3$ has been destroyed in the inner disk, relative to the primordial ice reservoir. 

\subsection{Nitrogen in outer disks and the cold interstellar medium}

In contrast to the inner disk, the outer disk is characterized by the sequestration of bulk volatiles into ices, with only a small fraction reaching the gas-phase due to the action of non-thermal desorption mechanisms. Recently, gas-phase NH$_3$ was detected in the outer disk of TW Hya \citep[50--100\,au,][]{Salinas16}. The location of the detected NH$_3$ gas within the TW Hya disk is not well constrained, and it may not be cospatial with the detected cold water vapor \citep{Hogerheijde11}. The detected gas has a low local abundance, but is thought to be a tracer of photo-evaporation, indicating the presence of a much more massive reservoir of ice. The inferred NH$_3$ ice abundance relative to water, although with significant uncertainty, could be higher than that of solar system comets by as much as an order of magnitude, thus potentially accounting for most of the elemental nitrogen. While the conversion between gas-phase abundances and the abundances of the underlying ice reservoir is uncertain, the current best evidence nevertheless suggests that NH$_3$ is an abundant carrier of nitrogen in the cold outer parts of protoplanetary disks.

In interstellar ices, NH$_3$ is the primary detected nitrogen carrier ($[{\rm NH_3}/{\rm H_2O}]\sim 0.1$) \citep{Bottinelli10,Oberg11}. The abundance of NH$_3$ in interstellar ices is even somewhat higher than in comets \citep{Oberg11}. While HCN is commonly detected in inner disks from mid-infrared spectroscopy \citep{Salyk11,Mandell12,Najita13}, and in outer disks through its rotational lines \citep{Oberg10}, it has yet to be detected in interstellar ices. 

\subsection{This paper}

With this paper, we aim is to derive a more comprehensive picture of the evolution of dominant nitrogen carriers toward the formation of planets. To this end, we employ new and archival infrared spectroscopy of warm gas in inner protoplanetary disks to constrain the major reservoirs of nitrogen near 1\,au. Specifically, we report a deep 10.7\,$\mu$m search, at high spectral resolution, for warm NH$_3$ at $\sim$1\,au in protoplanetary disks around solar-mass young stars. Using two-dimensional radiative transfer models to retrieve the amount of warm NH$_3$ in the disk surfaces, we derive robust upper limits on the NH$_3$ abundance, improving on previous limits on warm NH$_3$ gas in disks by up to an order of magnitude. We use the upper limits on inner disk NH$_3$ to investigate whether inner disk chemistry increases the average volatility of bulk nitrogen by destroying NH$_3$, and potentially re-partitioning it into highly volatile N$_2$. We also report new upper limits on the abundance of HCN ice in prestellar dust using archival M-band spectroscopy. Finally, we discuss the implications this may have on our understanding of delivery of nitrogen to the Earth and terrestrial exoplanets. 

\section{Observations and data}

\begin{figure*}[ht!]
\centering
\includegraphics[width=18cm]{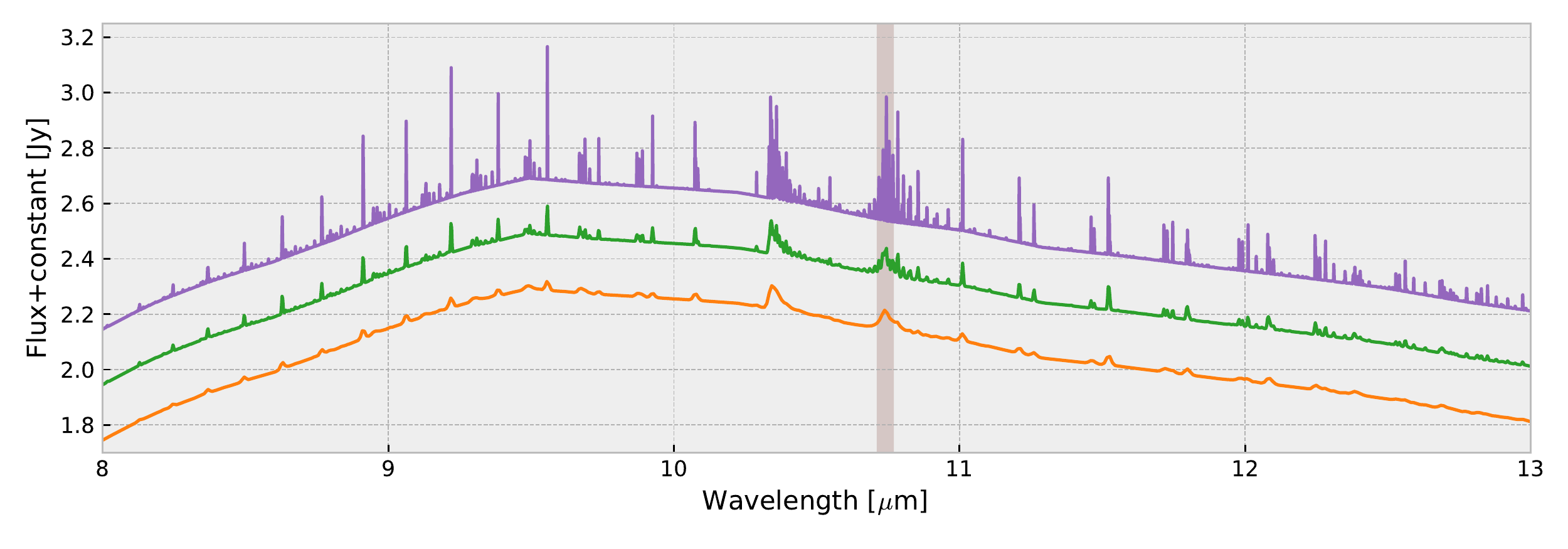}
\caption{Model spectrum showing the NH$_3$ $\nu_2$ band region at different spectral resolving powers, relevant for TEXES (top curve; $R=85\,000$), JWST-MIRI (middle curve; $R=2\,000$) and Spitzer-IRS (bottom curve; $R=600$). The shaded region indicates the coverage of the TEXES NH$_3$ spectral setting. The model shown assumes an inner disk NH$_3$ abundance of $2.5\times 10^{-7}$ per H. Note that in actual disk spectra, other molecular species contribute additional strong lines, some of which will be blended with some of the NH$_3$ lines.}
\label{fig:texes_setting}
\end{figure*}

To compare the chemical composition of carriers of bulk nitrogen in inner protoplanetary disks to outer disks, protostars and molecular clouds, we obtained high-resolution spectroscopy of a section of the strong 10\,$\mu$m NH$_3$ $\nu_2$ band (``umbrella mode'') for a small sample of protoplanetary disks using the TEXES spectrometer \citep{Lacy02} mounted on the Gemini North telescope on Mauna Kea, Hawai'i. This mode is one of the strongest NH$_3$ bands, and is located in a clear part of the Earth's transmission spectrum without strong telluric absorption lines. It has previously been detected in absorption from the ground, using TEXES, toward a number of massive protostars \citep{Knez09,Barentine12}, with inferred abundances of $[{\rm NH_3/H_{nuc}}]\sim 5\times 10^{-7}$, as well as in Jupiter's atmosphere \citep{Fletcher14}. The strongest lines in the covered spectral range (10.715--10.77\,$\mu$m) trace energies of 1400--1800\,K above the ground state. These are similar to the upper level energies of the rotational water lines in the 15-30\,$\mu$m range traced by Spitzer, which are known to originate in $\sim 500\,$K gas \citep{Carr11,Salyk11}. That is, the mid-infrared NH$_3$ and water lines will trace the same gas reservoir, and the observations will provide a robust measurement of the [{\rm NH$_3$/H$_2$O}] ratio, with the option to infer a local [NH$_3$/H$_{\rm nuc}$] abundance using a canonical [H$_2$O/H$_{\rm nuc}$] abundance of $5\times 10^{-5}$. In Figure \ref{fig:texes_setting} the TEXES setting is shown in the context of the wider NH$_3$ spectrum. 

We targeted three bright protoplanetary disks around solar-mass young stars, AS 205N, DR Tau and RNO 90, which were selected based on the presence of strong mid-infrared rotational water line emission and the ro-vibrational HCN $\nu_3$ band at 14\,$\mu$m \citep{Salyk11} as observed with the Spitzer Space Telescope InfraRed Spectrograph \citep[IRS, ][]{Houck04}. We also retrieved Spitzer high-resolution spectra of the same sample of disks from \cite{Pontoppidan10} to derive consistent HCN abundances using the strong 14\,$\mu$m $\nu_2$ bending mode. All three disks have velocity-resolved ground-based 12.4\,$\mu$m spectroscopy of a few of the water lines \citep{Pontoppidan10b,Banzatti14}, demonstrating that the water emission originates from the inner disks. The presence of water emission allows for the derivation of meaningful relative abundances of NH$_3$ and HCN. Further, all three disks orbit stars of $0.8-1.0\,M_{\odot}$, and of ages 1-2\,Myr, implying that the three objects trace general properties of this specific demographic.

Finally, we use archival ice spectroscopy of a sample of unrelated protostars to estimate new upper limits on the general HCN abundance in star-forming ices prior to the formation of protoplanetary disks. While the NH$_3$ abundance in circumstellar ices was measured by Spitzer \citep{Bottinelli10}, HCN has not yet been detected in the solid phase, and few or no upper limits exist in the literature. Specifically, we estimate upper limits on the HCN ice abundance in a sample of young stars with Spitzer-based NH$_3$ ice detections using the 4.757\,$\mu$m (2102\,cm$^{-1}$) HC-N stretch in spectra obtained with the ISAAC instrument on the ESO Very Large Telescope, as part of the large program 164.I-0605 \citep{Dishoeck03}. 

\subsection{Target properties}
RNO 90 is a solar-mass \citep[1.5\,$M_{\odot}$;][]{Pontoppidan11} star surrounded by a classical protoplanetary disk. It is also one of the brightest disks known to be rich in molecular emission lines in the infrared, with double-peaked line profiles and spectro-astrometry demonstrating an origin in a Keplerian disk \citep{Pontoppidan08}. AS 205N is an optical component in a 1.3'' binary. It is somewhat more massive than the Sun, and is still accreting at a relatively high rate of $8\times 10^{-8}\,M_{\odot}\,\rm{yr}^{-1}$ \citep{Prato03, Salyk13}. It has ro-vibrational CO line profiles that have been interpreted as evidence of a disk-wind flow \citep{Pontoppidan11, Salyk14}. Thus, AS 205N represents a more active, possibly younger system. DR Tau is a well-known variable, high-accretion T Tauri disk, and the brightest known molecule-rich disk in the Taurus star-forming region \citep{Carr11}. Similar to AS 205N, the CO profiles of DR Tau indicate a partial origin in a disk wind. 

\begin{deluxetable*}{lccccccc}[ht]
\tablecolumns{8}
\tablewidth{0pc}
\tablecaption{Source properties and TEXES observing log}
\tablehead{
\colhead{Disk} & \colhead{$M_{\rm *}$}   & \colhead{Luminosity}\tablenotemark{a} & \colhead{Distance} & \colhead{Incl.}     & \colhead{Date} & \colhead{Calibrator} & \colhead{Int. time}\\
\colhead{}     & \colhead{[$M_{\odot}$]} & \colhead{[$L_{\odot}$]}               & \colhead{[pc]}     & \colhead{[degrees]} &                &                      & \colhead{[Minutes]}}

\startdata
AS 205N        & 1.0\tablenotemark{b} & 4.1\tablenotemark{b} & 128\tablenotemark{d} & 20\tablenotemark{e} & 2014 Aug 11 & Eunomia & 19\\
DR Tau         & 0.8\tablenotemark{e} & 2.2\tablenotemark{c} & 196\tablenotemark{d} &  9\tablenotemark{e} & 2014 Aug 14 & Europa  & 26\\
RNO 90         & 1.5\tablenotemark{e} & 2.9\tablenotemark{c} & 117\tablenotemark{d} & 37\tablenotemark{e} & 2014 Aug 13 & Eunomia & 45
\enddata
\tablecomments{
\tablenotetext{a}{Total (stellar+accretion) system luminosity.}
\tablenotetext{b}{\cite{Andrews10}, corrected to the GAIA distance.}
\tablenotetext{c}{\cite{Blevins16}, corrected to the GAIA distance.}
\tablenotetext{d}{\cite{Gaia16,Gaia18}}
\tablenotetext{e}{\cite{Pontoppidan11}}
}
\label{table:source_properties}
\end{deluxetable*}

\subsection{TEXES spectroscopy of NH$_3$}
We obtained spectra of the $^{14}$NH$_3$ Q-branch of the $\nu_2$ band near 10.75\,$\mu$m (930\,cm$^{-1}$) toward the three protoplanetary disks around the solar-mass young stars AS 205N, RNO 90 and DR Tau. We used the cross-dispersed high-medium configuration, which results in instantaneous coverage between 10.715 and 10.775\,$\mu$m at a spectral resolving power of $\lambda/\Delta\lambda\sim 85,000$. A two-dimensional radiative transfer model of the RNO 90 disk \citep{Blevins16} predicts that this spectral region contains some of the strongest NH$_3$ lines visible from the ground (see Figure \ref{fig:texes_setting}).

The TEXES data were reduced using standard procedures, including division by a flat field, linearization, registration and co-addition of individual nod-cycles. The spectra were calibrated using equivalent observations of bright asteroids, which are used to remove the combined signature of the system spectral response function and telluric absorption. The wavelength solution was determined using an atmospheric model of the sky emission spectrum. Based on the fit of the observed sky background and the sky emission model, we estimate that the wavelength solution is accurate to $\sim 3\,{\rm km\,s^{-1}}$. 

\begin{figure*}[ht!]
\centering
\includegraphics[width=18cm]{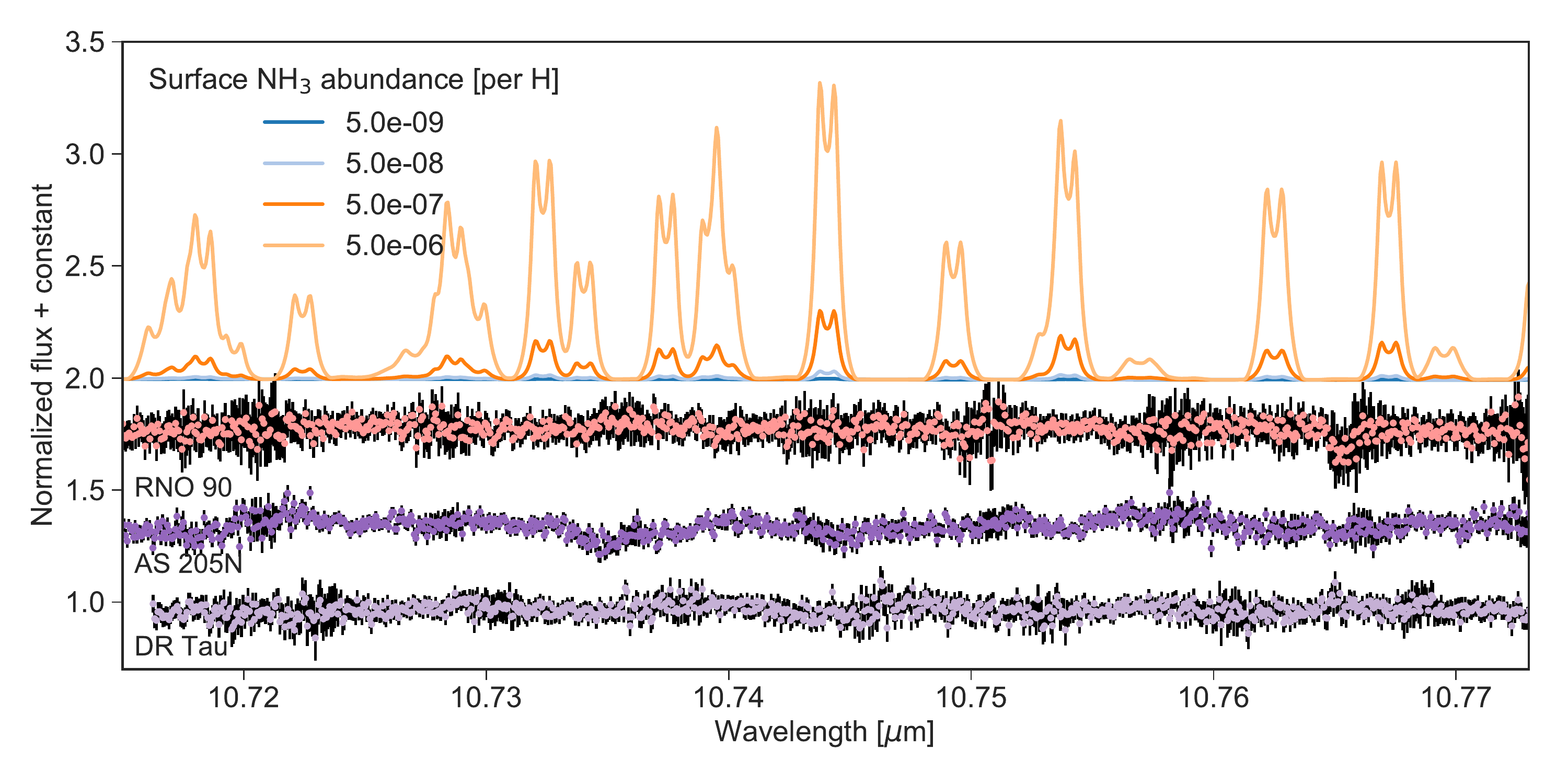}
\caption{High resolution TEXES spectra of part of the NH$_3$ $\nu_2$ mode, compared to the RNO 90 reference disk model \citep{Blevins16}, assuming different inner disk NH$_3$ abundances relative to H at $\sim$1\,au. Some response function and/or fringe residuals are seen, but these are not coincident with the expected NH$_3$ lines, and tend to be much broader. The derived upper limits fall between the models with NH$_3$ abundances of $5\times 10^{-8}$ and $5\times 10^{-7}$ per H.}
\label{fig:texes_data}
\end{figure*}

The reduced TEXES spectra are presented in Figure \ref{fig:texes_data}. They show no detection of NH$_3$ lines in any of the disks down to the 5-10\% line-to-continuum level. Some residual low-frequency noise is apparent, which may be due to residual fringing not removed by the telluric calibrators. This leads to spurious features with widths similar to those of individual spectral orders ($\sim 0.01\,\mu$m, or $200-300\rm\,km\,s^{-1}$). However, based on the observed widths of the (energetically similar) CO fundamental ro-vibrational lines of $37-92\rm\,km\,s^{-1}$ \citep{Banzatti15}, the low-frequency noise is not likely to be confused with intrinsic NH$_3$ line emission. 

\subsection{Archival observations of HCN ice in circumstellar envelopes}
In addition to NH$_3$, HCN is another potential carrier of significant amounts of nitrogen. While there are many detections of HCN gas in protoplanetary disks \citep{Dutrey97,Oberg11b}, as well as in protostellar envelopes \citep{Jorgensen04}, there is, to our knowledge, no direct detection of a solid counterpart in interstellar ices, nor have any upper limits been reported. Indirect tracers of solid HCN exist, such as observations of warm gas in hot core regions, thought to trace recently evaporated ices. Based on inference from observations of warm gas-phase HCN, HCN ice is expected to have relatively low abundance compared to major ice species; approximately $2.5\times 10^{-7}$ relative to H, or 0.3-0.5\% relative to water \citep{Lahuis00}. 

In order to better estimate the primordial HCN content of primordial ices, for comparison with the observed abundances of HCN in disks, we report estimated upper limits on HCN ice on grain mantles present during the cold phases of star formation. We use archival M-band spectra from Keck-NIRSPEC and VLT-ISAAC of the strong, narrow 4.75\,$\mu$m (2100\,cm$^{-1}$) C-N stretch of HCN toward a number of circumstellar envelopes around low-mass young stars. This is the most promising band for HCN detection, as the stronger C-H stretch around 3.2\,$\mu$m is blended with the 3.1\,$\mu$m water ice band, and other bands are weaker \citep{Gerakines04}. The ISAAC spectra were previously published in \cite{Pontoppidan03}, whereas the Keck-NIRSpec spectra of RNO 91 and EC 90 have not previously been published. We include sources from \cite{Pontoppidan03} for which the water ice band has an optical depth $\tau>1$, and with a signal-to-noise ratio of $\sim$50 or more.

\begin{figure}[ht!]
\centering
\includegraphics[width=8cm]{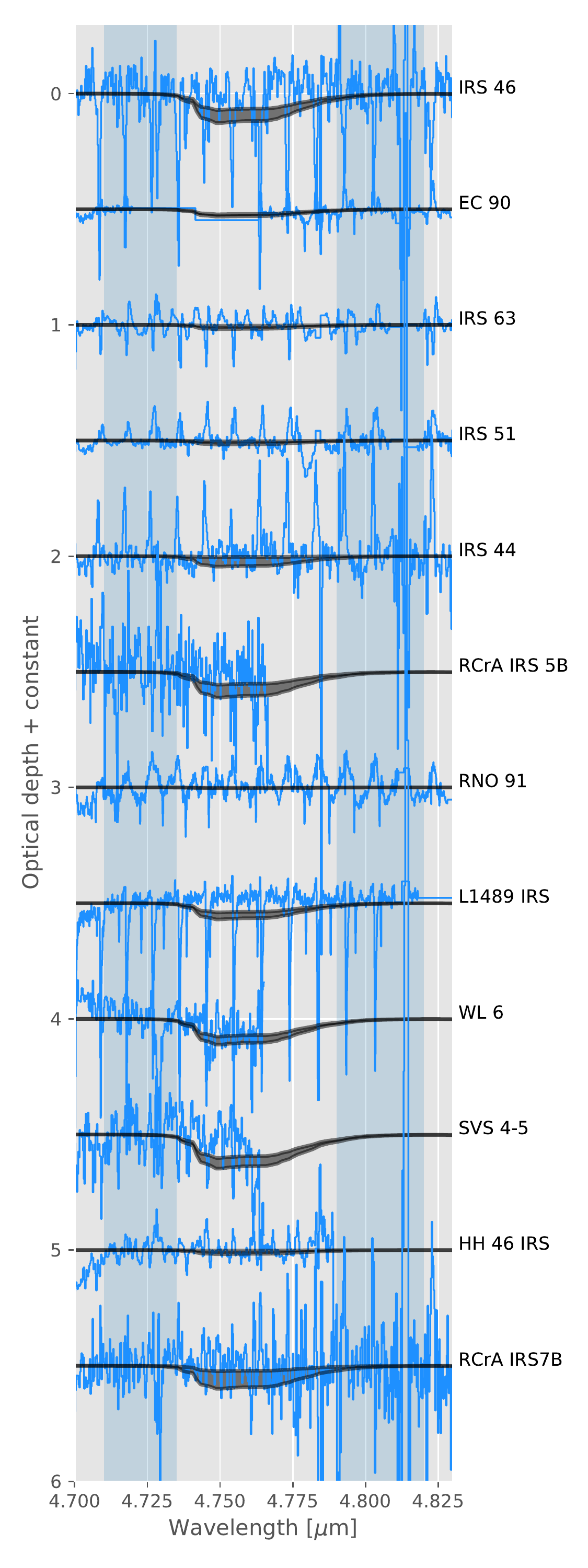}
\caption{Fits of the 4.75\,$\mu$m (2100\,cm$^{-1}$) band of solid HCN to archival M-band spectra of nearby embedded young stars. The curves show the 1 and 5$\sigma$ upper limits on the strengths of the HCN band. The vertical rectangles indicate the regions of the spectrum used to define the continuum. The deep absorption feature at 4.67\,$\mu$m is due to CO ice. Also visible in most spectra are emission and absorption lines from ro-vibrational CO gas-phase transitions.}
\label{fig:hcn_abundance}
\end{figure}

A challenge to using the CN stretch to detect HCN ice is that it is often affected by absorption and/or emission from warm gas-phase CO, requiring high resolution spectroscopy ($R\gtrsim 10\,000$) to clearly separate any broad ice absorption from multiple CO gas lines. We define the continuum as a linear function, fit to regions between 4.71--4.735\,$\mu$m and 4.79--4.82\,$\mu$m. The optical depth spectra are then calculated as $\tau = -{\mathrm ln}(F/C)$, where $F$ and $C$ are the flux density and continuum spectra, respectively. The upper limits are derived as $3\sigma$ values using the pure HCN ice spectrum and band strength ($5.1\times 10^{-18}\,\rm cm\,molec^{-1}$) from \cite{Gerakines04}. The shape of the HCN band is somewhat sensitive to composition of the ice mantles, with mixtures with other polar molecules, such as water, leading to signifant broadening and reddening \citep{Noble13}. The resulting upper limits on the HCN ice column densities are given in Table \ref{hcnice}. Compared to the water ice column densities toward the same sources, the upper limits on HCN generally correspond to maximum [HCN/H$_2$O] abundances of 2-5\%, consistent with the relative abundance of gas-phase HCN observed in hot cores. It follows that improvements in sensitivity to solid HCN by at least an order of magnitude are likely necessary to detect it in protostellar sightlines. 

\begin{deluxetable*}{lccccc}[ht]
\tablecolumns{6}
\tablewidth{0pc}
\tablecaption{Nitrogen-bearing ice column densities in protostellar envelopes}
\tablehead{
\colhead{Disk} & \colhead{$N({\rm NH_3})$} & \colhead{$N({\rm H_2O})$} & \colhead{$N({\rm HCN})$} & \colhead{Instrument} & \colhead{Reference}
}
\startdata
IRAS 08242-5050 & $4.77\pm0.46$ & $77.9\pm7.7$ & $<2.7$ & VLT-ISAAC & \citep{Pontoppidan03} \\
SVS 4-5 & $2.4:$ & $56.5\pm11.3$ & $<4.0$ & VLT-ISAAC & \citep{Pontoppidan03} \\
R CrA IRS5 & $0.91\pm0.23$ & $35.8\pm2.6$ & $<2.6$ & VLT-ISAAC & \citep{Pontoppidan03} \\
L1489 IRS & $2.31\pm0.3$ & $42.6\pm5.1$ & $<1.7$ & Keck-NIRSPEC & \citep{Boogert02} \\
RNO 91 & $2.03\pm0.3$ & $42.5\pm3.6$ & $<1.5:$ & Keck-NIRSPEC & This paper \\
EC 90 & $0.67\pm0.2$ & $16.9\pm1.6$ & $<0.8$ & Keck-NIRSPEC & This paper \\
CrA IRS 7B & $3:$ & $110.1\pm19.7$ & $<1.9$ & VLT-ISAAC & \citep{Pontoppidan03} \\
WL 6 & $1.2093\pm0.24$ & $41.7\pm6$ & $<2.9$ & VLT-ISAAC & \citep{Pontoppidan03} \\
IRS 44 & $1.258\pm0.25$ & $34\pm4$ & $<1.2$ & VLT-ISAAC & \citep{Pontoppidan03} \\
IRS 46 & $0.65\pm0.13$ & $12.8\pm2$ & $<3.1$ & VLT-ISAAC & \citep{Pontoppidan03} \\
IRS 51 & $0.53\pm0.11$ & $22.1\pm3$ & $<0.3$ & VLT-ISAAC & \citep{Pontoppidan03} \\
IRS 63 & $1.16\pm0.23$ & $20.4\pm3$ & $<0.3$ & VLT-ISAAC & \citep{Pontoppidan03} 
\enddata
\tablecomments{All column densities are provided in units of $\rm 10^{17}\,cm^{-2}$. Colons indicate high uncertainty. 
}
\label{hcnice}
\end{deluxetable*}

\section{Analysis}
\subsection{Radiative transfer modeling and abundance retrieval}
In order to retrieve upper limits of the local abundance of NH$_3$ gas at $\sim 1$\,au, we use the modeling framework described in \cite{Zhang13} and \cite{Blevins16}. This framework uses the two-dimensional line radiative transfer code RADLite \citep{Pontoppidan09} to render predicted line spectra given a disk density structure and molecular abundance structure. The model was fitted in detail to the broad-band SED and 10-180\,$\mu$m Spitzer-IRS and Herschel-PACS water spectra for RNO 90. It was used to retrieve inner disk H$_2$O and CO abundances \citep[see ][for details on the modeling approach]{Pontoppidan14,Blevins16}. 

The model includes both dust and gas. The dust density structure is fitted to the observed continuum Spectral Energy Distribution (SED). The dust model includes dust grains up to 40\,$\mu$m in size, with a power law size index of -2.5, relevant for the disk surface traced by the infrared molecular lines; larger grains are assumed to have settled to the midplane, where they no longer contribute to the infrared properties of the disk. The molecular abundances are degenerate with respect to the assumed gas-to-dust ratio, such that lower absolute abundances tend to be needed to produce a given observed line flux for higher gas-to-dust ratios. This is due to a larger gas column being visible for lower opacity disks. Following the procedure in \cite{Blevins16}, we assume a constant gas-to-dust mass ratio of 100 everywhere in the disk. While this is unlikely to be universally true, the infrared molecular spectra trace a relatively small part of the disk (the surface at $\sim$1 au), so the assumed gas-to-dust ratio is relevant for a relatively restricted region. There is significant uncertainty in the surface gas-to-dust ratio as competing processes are thought to be active. For instance, dust settling and growth acts to increase the gas-to-dust ratio \citep{Meijerink09, Horne12, Carmona14}, disk dispersal mechanisms such as photoevaporation tend to decrease it \citep{Bruderer14, Ansdell16} and chemical effects may mask the true ratio \citep{Kama16,Miotello17}. 

The gas temperature is estimated using the thermo-chemical calculation of \cite{Najita11}, and scaled to the specific RADLite dust model using the vertical column density at each disk radius (see Figure 4 in \cite{Blevins16}). The level populations are assumed to be in local thermodynamic equilibrium. While the critical densities of infrared molecular transitions for collisions with H$_2$ tend to be high ($n\sim 10^{10}-10^{12}\,\rm cm^{-3}$), a combination of infrared pumping from the local warm dust and with collisions with atomic H provides an efficient mechanism leading to near-thermalization. This has been observed both for rovibrational CO lines \citep{Blake04, Thi13}, as well as for the mid-infrared rotational H$_2$O lines (with high rotational quantum numbers) \citep{Meijerink09}, and has been supported by models for HCN \citep{Bruderer15}. For disks around early-type stars, UV fluorescence can in some cases also be important \citep{Brittain07}, but this is expected to be a small effect for disks around low-mass and solar-mass stars where the local UV fields are less intense. While the expectation is that non-LTE effects are relatively small, in particular for lower-mass stars with low UV fields, non-LTE models for NH$_3$ and HCN are ultimately needed to confirm the results presented in this work.

\begin{figure*}[ht!]
\centering
\includegraphics[width=8.7cm]{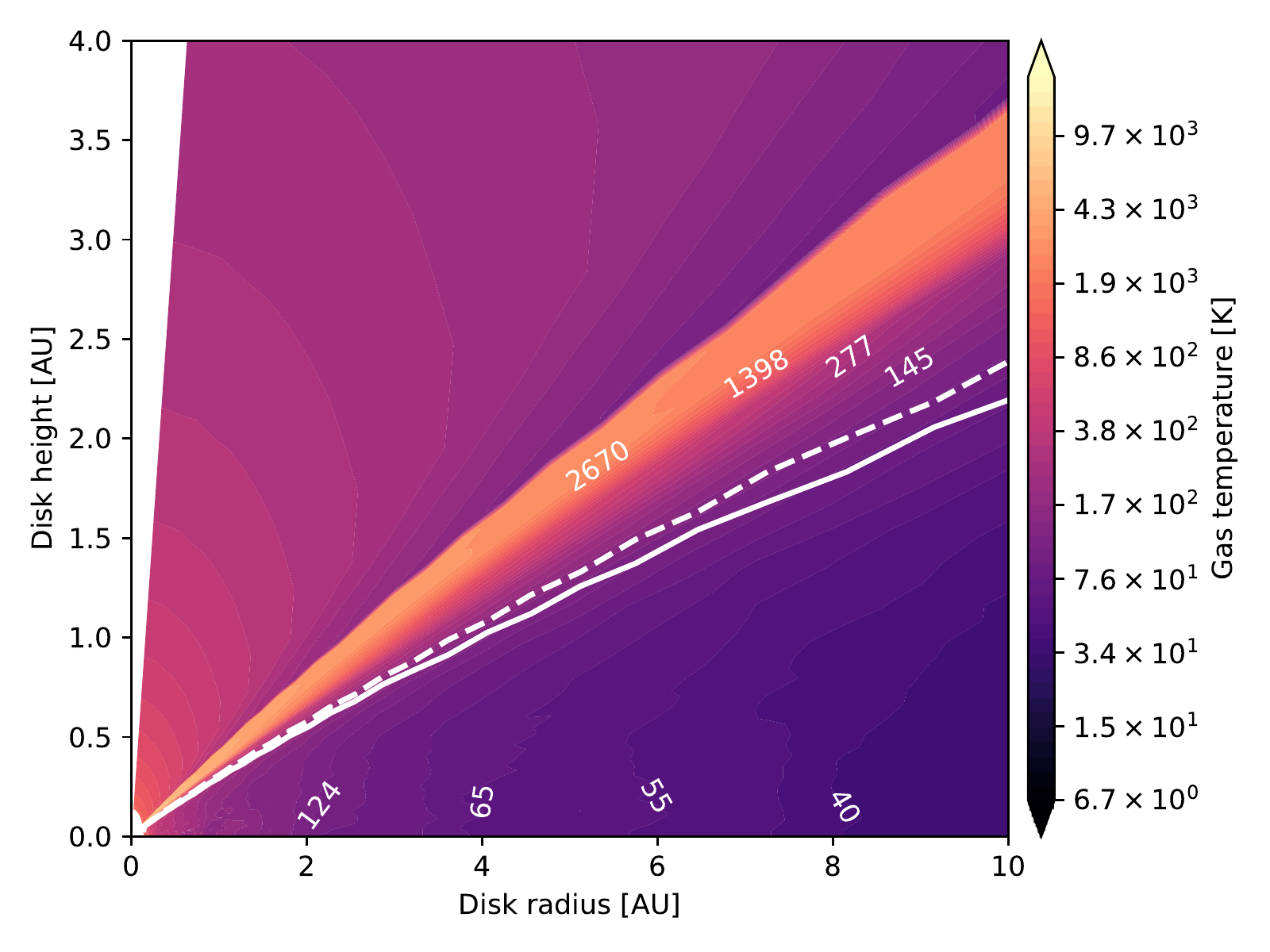}
\includegraphics[width=8.7cm]{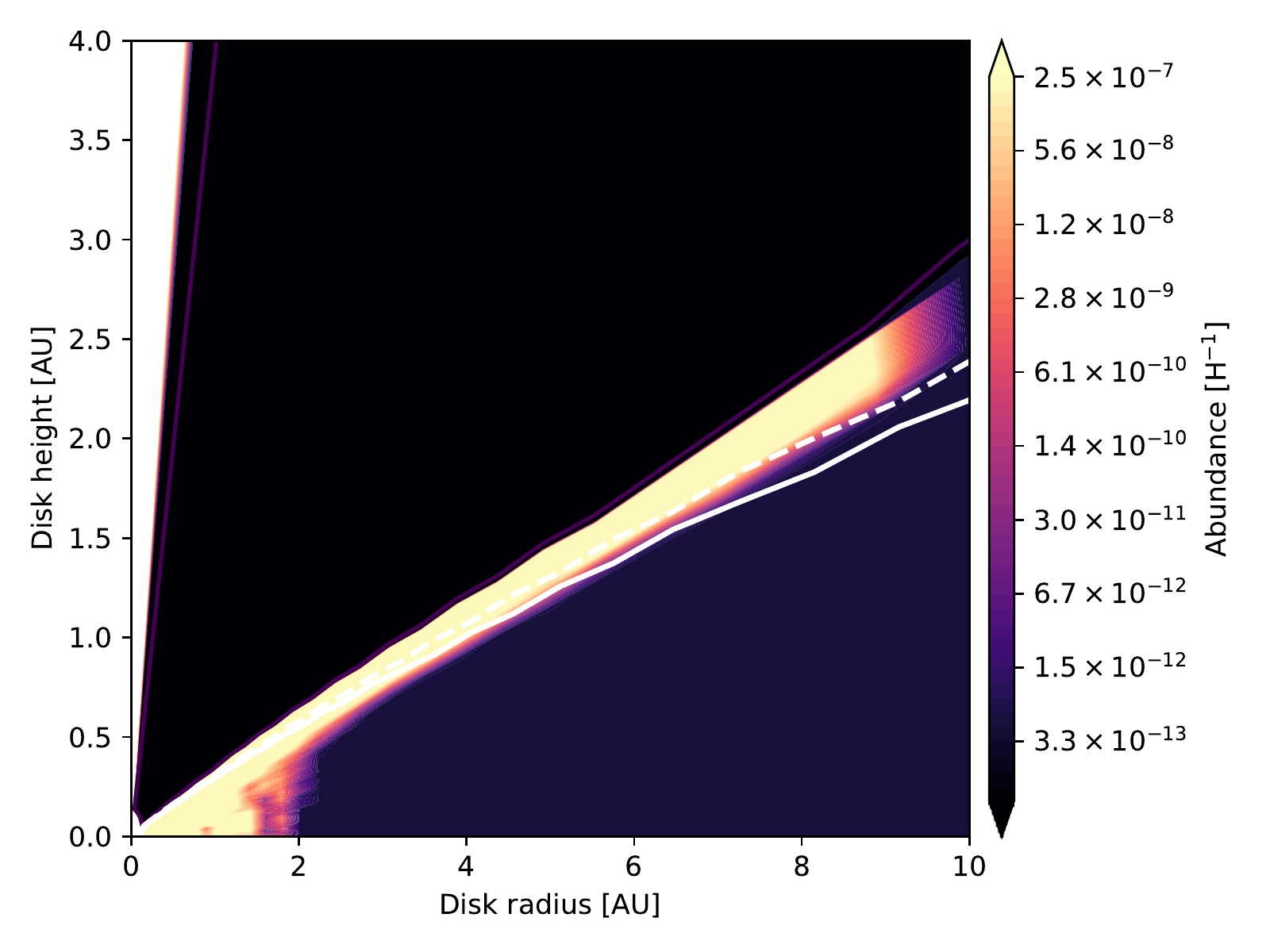}
\caption{Left: the gas temperature structure. Some temperature contours are indicated in units of K. Right: The NH$_3$ abundance of a model close to the upper limit. The solid white curves show the $\tau=1$ surface at 10.7\,$\mu$m, whereas the dashed curves indicate the surface corresponding to a vertical column density of $10^{22}\,\rm cm^{-2}$.}
\label{fig:model_abundance}
\end{figure*}

The warm water and CO ro-vibrational line-to-continuum ratios in all three disks fall within a factor of 2-3 of each other, and the previous applications of this retrieval approach indicate water and CO abundances that are consistent with canonical abundances of $[{\rm H_2O/H_{nuc}}] \sim [{\rm CO/H_{nuc}}] \sim 5\times 10^{-5}$ \citep{Pontoppidan14b}. The three sources also have similar spectral energy distributions, and their luminosities all fall within a factor 2 (see Table \ref{table:source_properties}). Consequently, we consider the three disks to differ only to first order, which means we can use the model spectra for RNO 90, scaling by constant factors to the observed 10.7\,$\mu$m continua of AS 205N and DR Tau. The implicit assumption is that, all other model parameters (abundance structure) being equal, infrared line fluxes scale linearly with source luminosity within the narrow range of luminosity of the sample (as does the continuum). Even if there are inherent uncertainties in the absolute abundance relative to H, the abundances relative to water and CO will be self-consistent. 

The molecular abundances are implemented as ``jump models'', in which the abundance is high inside a critical radius, $R_{\rm crit}$, and low outside. Further, the gas abundance is assumed to be low where the dust temperature is below the relevant freeze-out temperature. If $R_{\rm crit}$ is very large, the disk abundance structure follows the location of the snow line. Conversely, if $R_{\rm crit}$ is small, the abundance jump may be interpreted as being due to a chemical effect, rather than freeze-out. 

\begin{figure}[ht!]
\centering
\includegraphics[width=8.5cm]{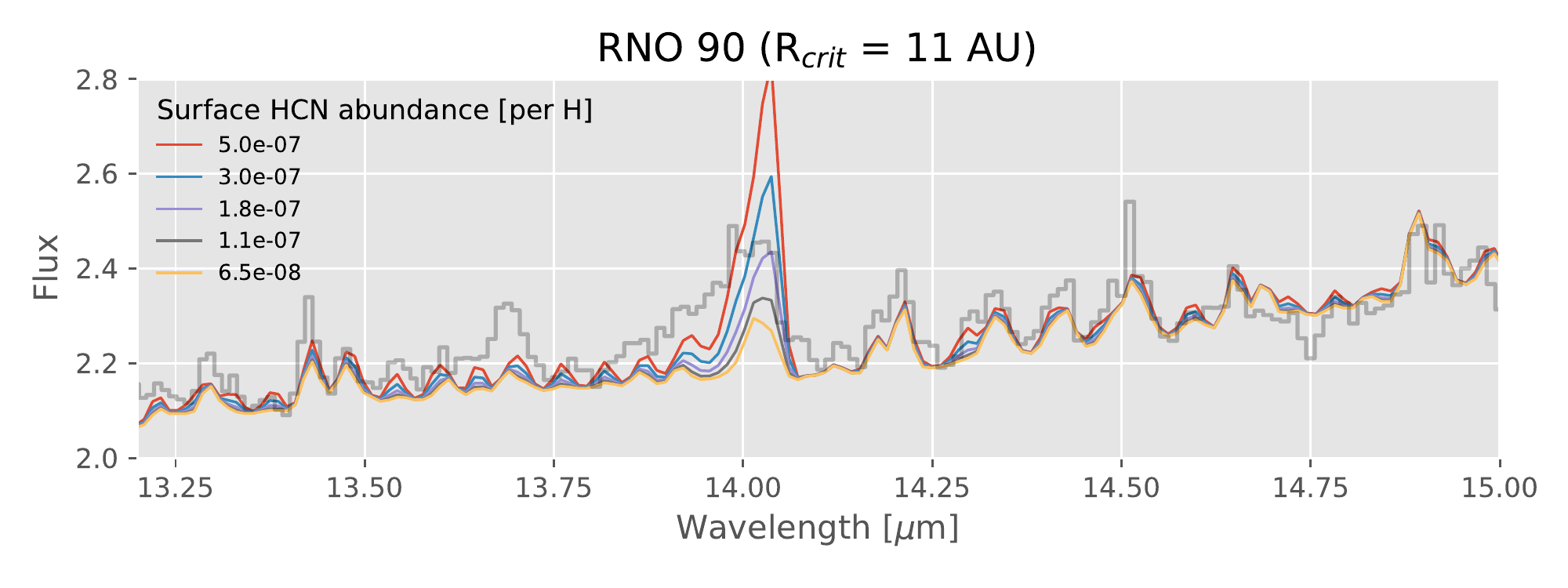}
\includegraphics[width=8.5cm]{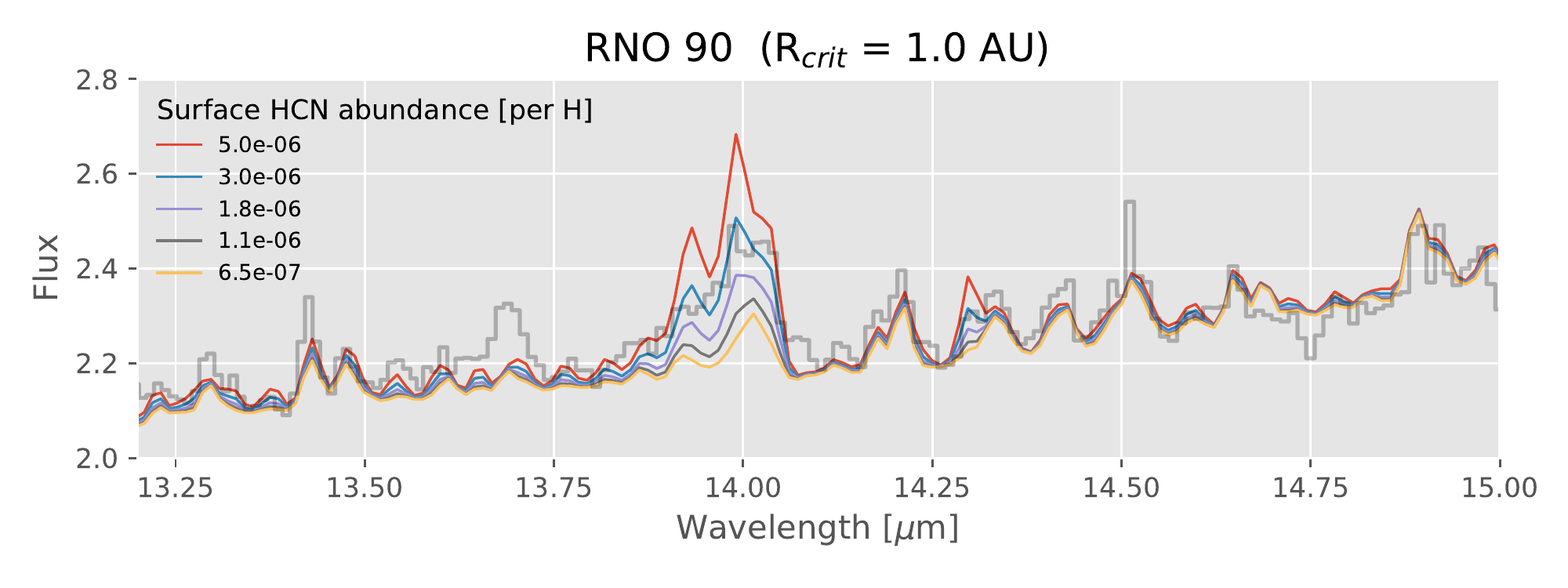}
\includegraphics[width=8.5cm]{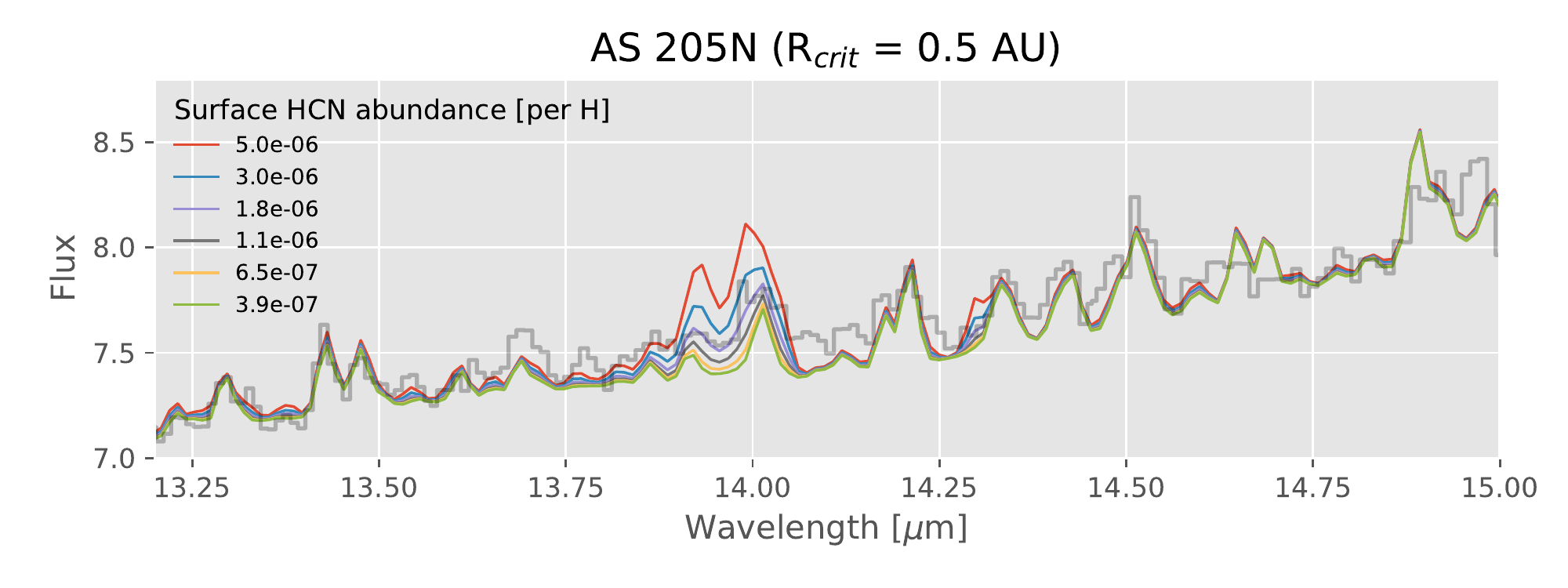}
\includegraphics[width=8.5cm]{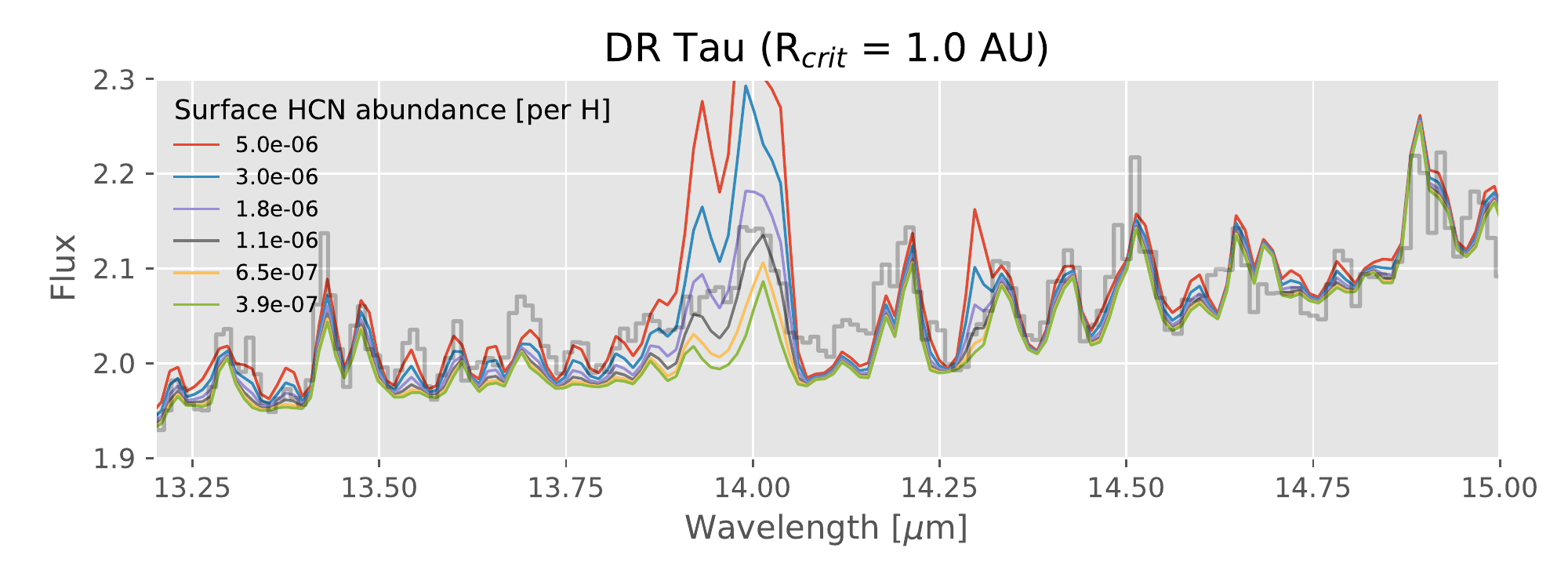}
\caption{Fits of the inner disk HCN abundance to archival Spitzer spectra using the RADLite grid.  The water abundance is held constant at [H$_2$O/H$_{\rm nuc}$]=$5\times 10^{-5}$, while the HCN abundance and radial size of the region where the HCN abundance is high are varied. The RNO90 fit is consistent with that presented in \cite{Pontoppidan14}. Top: Best fits assuming a large critical radius. Bottom three: Best fits assuming $R_{\rm crit} \leq 1\,$au.}
\label{fig:hcn_gas_abundance}
\end{figure}

\subsection{Inner disk NH$_3$ upper limits}
Using the RNO 90 structure model, we calculate a grid of models by varying the NH$_3$ gas abundance between $[{\rm NH_3/H_{nuc}}] = 5\times 10^{-8}$ and $10^{-6}$, as models outside of this range are clearly inconsistent with the data (see Figure \ref{fig:texes_data}). Since more face-on configurations will increase the line-to-continuum ratios, and therefore the detectability for the same abundance, we calculate a separate model grid for the known disk inclinations (see Table \ref{table:source_properties}). The non-detections offer no constraints on a critical radius, so we assume the NH$_3$ abundance has a large $R_{\rm crit}=11\,$au, which is the measured location of the surface water snowline in RNO 90 by \cite{Blevins16}. The mid-plane water snowline in the same disk is at 1\,au, but this is not likely to be traced by our data. For each model in the grid we then identify spectral channels of the model spectrum with significant potential line emission ($>10\%$ of the peak line flux) and calculate the least-squares ($\chi^2$) statistic relative to the observed spectrum on those channels. The model with the lowest abundance that can be excluded with 99\% confidence is identified and adopted as the appropriate upper limit. A caveat to upper limits is that we have no measurement of line ratios and opacities, and we therefore cannot rule out higher local NH$_3$ abundances with a small filling factor. That is, it is possible that high abundance NH$_3$ is concentrated in a small area ($\ll 10$\,AU); we can only say that the average abundance in the inner disk surface is low. 

\subsection{New inner disk HCN abundances}
Since a key comparative nitrogen reservoir in the inner disk is warm HCN, we use the same radiative transfer framework to retrieve the HCN abundance by fitting to the Spitzer spectra presented in \cite{Pontoppidan10}. \cite{Salyk11} estimated abundances for the same data set using a simple slab model, however a more self-consistent 2-dimensional RADLite model is expected to significantly improve the estimate. The HCN abundance is constrained not only by the fundamental 14\,$\mu$m $v=01^10\rightarrow 00^00$ intensity, but also by the shape of the band and the presence of the $v=02^00\rightarrow 01^10$ band near 14.3\,$\mu$m. These bands have different opacities and excitation temperatures, and allow for an independent estimate of the critical radius of a jump model. For HCN, we find that a small critical radius of $1\pm 0.2\,$au is needed to fit the shape of both the 14 and 14.3\,$\mu$m bands. This is illustrated in Figure \ref{fig:hcn_gas_abundance}, where a large radius leads to HCN with too low opacity, as seen by a narrow 14\,$\mu$m band and relatively weak 14.3\,$\mu$m band. 

\cite{Bruderer15} presented a non-LTE model for HCN in the AS 205N disk. They found that the primary excitation mechanism is radiative pumping, and that retrieved abundances using the 14\,$\mu$m band are essentially unaffected by an assumption of LTE. Their model also provides a benchmark to the RADLite model, as they estimate the HCN for the same Spitzer spectrum of the AS 205N disk. For AS205N, we estimate an absolute abundance of $\rm [HCN/H_{nuc}]=2\times 10^{-6}$ for a gas-to-dust ratio of 100. In comparison, \cite{Bruderer15} also find that a sharp jump is needed around 1\,au and estimate an inner abundance of $\rm [HCN/H_{nuc}]=3\times 10^{-7}$ for an assumed gas-to-dust ratio of 1000. These values are consistent given the degeneracy of the assumed gas-to-dust ratio (if Bruderer et al. had assumed a smaller gas-to-dust ratio, the retrieved abundance would increase in proportion). 

The NH$_3$ limits and HCN gas abundances are shown in Table \ref{table:limits}.

\begin{deluxetable*}{lcccc}
\tablecolumns{5}
\tablewidth{0pc}
\tablecaption{Observed gas-phase nitrogen carriers in inner disks}
\tablehead{
\colhead{Disk} & \colhead{[HCN/H$_{\rm nuc}$]} & \colhead{[NH$_3$/H$_{\rm nuc}$]\tablenotemark{a}} & \colhead{[H$_2$O/H]} & \colhead{\% Missing N\tablenotemark{b}}
} 
\startdata
AS 205N & $(2\pm 0.5)\times 10^{-6}$ & $<1.1\times 10^{-7}$ & $5\times 10^{-5}$ & 97.0 \\
DR Tau & $(1\pm 0.2)\times 10^{-6}$ & $<1.4\times 10^{-7}$ & $5\times 10^{-5}$  & 95.5 \\
RNO 90 & $(3\pm 1.0)\times 10^{-6}$ & $<2.5\times 10^{-7}$ & $5\times 10^{-5}$ & 98.5
\enddata
\tablecomments{All absolute abundances are referenced to an assumed gas-to-dust ratio of 100.
\tablenotetext{a}{Upper limits are $3\sigma$}
\tablenotetext{b}{Assuming a solar nitrogen abundance of $[{\rm N/H_{nuc}}] = 6.76\times 10^{-5}$ \citep{asplund09}.}
}
\label{table:limits}
\end{deluxetable*}

\section{Discussion}

\subsection{Comparison to NH$_3$ in the outer disk}
A detection of high abundances of cold NH$_3$ gas was reported by \cite{Salinas16} using Herschel-HIFI observations of the ortho-NH$_3$ $1_0-0_0$ line at 572.5\,GHz. The line was reported to have a spectral FWHM of 0.9\,$\rm km\,s^{-1}$, which corresponds to disk radii $\gtrsim 30$\,AU, assuming a stellar mass of 0.7\,$M_{\odot}$ \citep{Herczeg14} and an inclination of the outer disk of $7^{\circ}$ \citep{Qi04}. While the retrieved gas-phase NH$_3$ abundance is low relative to the total nitrogen content, its presence is interpreted as being the photo-desorption product of a much larger reservoir of NH$_3$ ice. The inferred nebular ice abundance, while uncertain, could be as high as $[{\rm NH}_3/{\rm H}] \sim 10^{-5}$, even higher than that inferred by solar system comets (up to $[{\rm NH}_3/{\rm H}] \sim 10^{-6}$; see references in Figure \ref{fig:abundance_overview}). The inferred total NH$_3$ abundance in the outer disk of TW Hya is comparable to that of the most enriched interstellar ices observed in young stellar envelopes \citep{Bottinelli10}. 

\subsection{Constraints on the nitrogen budget}
In combination with existing detections of warm HCN from the same region of the disk, the presented NH$_3$ observations constrain the budget of elemental nitrogen, not sequestered in N$_2$, in the inner disk surface. We find that at radii inside of $10\,$au, the average NH$_3$ abundance is at least an order of magnitude lower than that of interstellar ice, and at least marginally lower than that of comets. We rule out the possibility that a significant amount of nitrogen is sequestered in NH$_3$ in the inner disk surface. In comparison, the inner disk HCN abundance is higher than that of comets at 1\,au, but likely significantly lower at radii between 1 and 10\,au. This indicates that the chemistry of disk volatiles in the terrestrial region has been significantly altered from its primordial state, possibly with a greater fraction of the nitrogen driven into N$_2$. 

In Figure \ref{fig:abundance_overview}, we compare the observed abundances, relative to water, of HCN and NH$_3$ in inner regions of disks to those of comets and interstellar ices. The figure demonstrates that the new upper limits on the inner disk NH$_3$ abundance are significantly lower than those observed in reservoirs tracing colder material. Although the NH$_3$ abundance in comets is already known to be depleted relative to interstellar ices, the inner disk abundance appears to be depleted by at least another order of magnitude. At the same time, a self-consistent abundance retrieval suggests that HCN abundances are somewhat enhanced in the inner disk. The upper limits on HCN in interstellar ices are not constraining, as they are consistent with both comets and the inner disk abundances. The implication is that some of the nitrogen lost from NH$_3$ could be driven into HCN in the innermost disk, inside of 1\,au. Outside of this region, the HCN abundance remains low. The presented data do not reveal where the liberated nitrogen went in the $\gtrsim 1\,$AU region, but chemical models suggest N$_2$ is a likely candidate (see Section \ref{section:chem_models}).

An important caveat is that the observations only probe the disk surface, down to vertically integrated column densities of a few $\times 10^{22}\,\rm cm^{-2}$ (see Figure \ref{fig:model_abundance}). Thus, the mid-plane region, where the bulk of the molecular mass resides is not directly constrained. Nevertheless, given that all disk material is inherited from a protostellar envelope, the fact that the observed upper limits on the NH$_3$ abundance in the inner disk are $\sim$50 times lower than that observed in the outer disk of TW Hya is strong evidence for the efficient destruction of NH$_3$ in inner disk surfaces. 

\subsection{Comparison to chemical models}
\label{section:chem_models}
In recent years, a number of studies of the observable chemistry of inner disks have become available to better understand the observed infrared tracers of warm molecular gas in disks \citep{Heinzeller11, Schwarz14, Walsh15, Agundez18}. Models including full nitrogen-bearing chemical networks predict that the bulk carriers of nitrogen in protoplanetary disks consist of N$_2$, HCN and NH$_3$, with only relatively minor contributions from other simple species. Disk models by \cite{Schwarz14} predict large abundances of gas-phase NH$_3$ at a few au in the midplane exceeding $10^{-5}$ in models that initialized with most nitrogen in NH$_3$, and an order of magnitude less in models initialized with most nitrogen in N or N$_2$. Disk models also generally predict large abundances of NH$_3$ ice throughout the disk midplane at radii where the dust temperature remains below $\sim 60\,$K, and that there is a tendency to form more NH$_3$ ice with time \citep{Furuya14}. 

If photodissociation is the main destruction mechanism of all three species, which is an appropriate assumption within a column density of $10^{22}\,\rm cm^{-2}$, the following destructive reactions apply:

\begin{align}
\mbox{N}_2 + h\nu &\rightarrow \mbox{N} + \mbox{N} \\ 
\mbox{HCN} + h\nu &\rightarrow \mbox{H} + \mbox{CN} \\ 
\mbox{NH}_3 + h\nu &\rightarrow \mbox{NH}_2 + \mbox{H} \mbox{ or } \mbox{NH} + \mbox{H}_2 
\end{align}

For a 4000\,K blackbody, the unshielded photodissociation rates for the above three reactions are: $3.2\times 10^{-16}\,\rm s^{-1}$, $5.7\times 10^{-12}\,\rm s^{-1}$, and $3.6\times 10^{-9}\,\rm s^{-1}$, respectively \citep{Heays17}. The balance of N between these three species will be governed by the rates of destruction versus reformation from constituent atoms/radicals. The unshielded destruction rates already indicate a preferential depletion of NH$_3$ and HCN relative to N$_2$ in the inner disk surface. Furthermore, we have recently learned that N$_2$ self-shields efficiently \citep{Li13,Heays14}. N$_2$ is therefore even more robust against photodissociation than both HCN and NH$_3$ in the disk surface. Together, this provides theoretical support for the preferential sequestration of N from NH$_3$ into N$_2$. Reformation of HCN will occur through CN + H$_2$ which has a small barrier easily overcome in warm-to-hot gas.  Reformation of NH$_3$ from N, NH, or NH$_2$, on the other hand, has several barriers, which will slow this reaction sequence down, even in warm-to-hot gas.  Any available atomic N is easily converted to N$_2$ via  barrierless reactions, e.g., $\rm N + NO\rightarrow N_2+O$ or $\rm N + CN\rightarrow N_2+N$, provided there is adequate supply of small radicals in the gas-phase. 

The discrepancy between the low NH$_3$ abundance found in the inner disk and the potentially large reservoir of NH$_3$ ice in the outer disk is that, once the temperature is sufficiently cold to retain both N, NH,  NH$_2$, and atomic H on dust grain surfaces, then the formation of NH$_3$ ice via sequential hydrogenation is rapid. Once formed, the high binding energy of NH$_3$ protects it in the ice until the dust grains are warmed to the sublimation temperature, at which point gas-phase chemistry takes over. Indeed, HCN, has an experimental binding energy on astrophysical surfaces in the range 3400-3800\,K with some uncertainty \citep{Noble13,Rice18}, whereas NH$_3$ has a binding energy of 4000\,K on compact water ice. When mixed with water NH$_3$ desorbs at the same temperature as the water, or $\sim 145\,K$ at laboratory pressures \citep{He16}. Chemical models therefore predict a reservoir of abundant gas-phase NH$_3$ within its snowline in the disk midplane due to a lack of efficient destruction mechanisms deeper in the disk. In this regime, NH$_3$ is mostly chemically inert, although cosmic rays can modify the composition and elemental partitioning over long timescales \citep[$>1$\,Myr; see][]{Eistrup16,Eistrup18}. Over time, cosmic ray processing also has the effect of driving NH$_3$ into N$_2$ \citep{Eistrup18}. 

Ultimately, the ability of the inner disk to retain abundant NH$_3$ likely depends on the efficiency of vertical mixing to expose shielded NH$_3$ from the midplane to the surface chemistry, and mix the processed results back down. Both vertical and radial transport are known to link disparate chemical regions in the disk \citep{Ilgner04, Semenov10}. More specifically, the theoretical expectation is that vertical mixing will transport chemically rich material upwards, and increase the abundances of many species in the inner disk surface, including NH$_3$ \citep{Heinzeller11}. The fact that we do not see abundant NH$_3$ in the surface could mean that either vertical mixing is weak, or that the surface destruction of NH$_3$ is fast. An argument against weak vertical mixing is that the water abundance is known to be high in all three observed disks, suggesting that we are indeed witnessing the active destruction of NH$_3$ and transfer of nitrogen into N$_2$. 

\begin{figure*}[ht!]
\centering
\includegraphics[width=16cm]{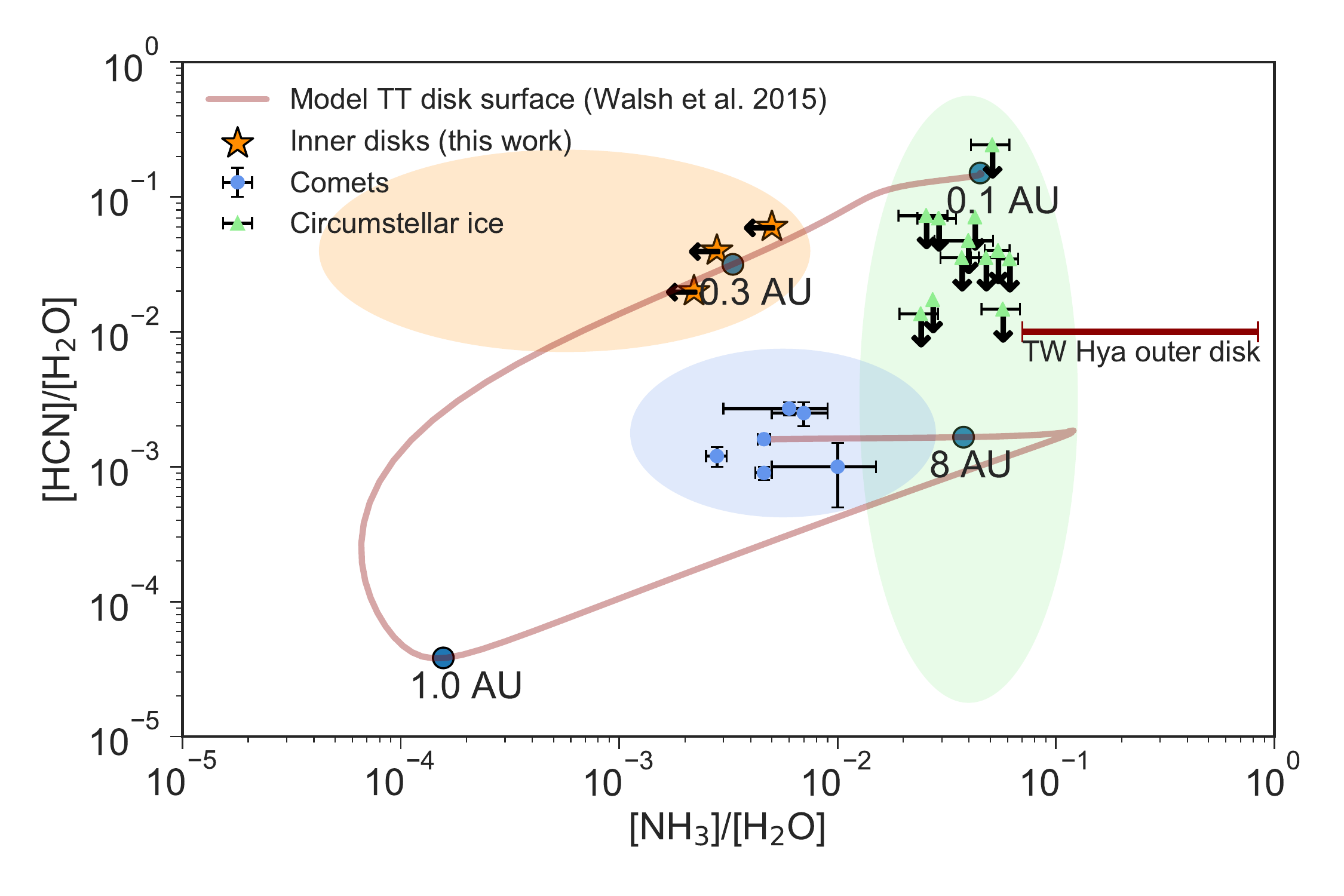}
\caption{Observed evolution of the nitrogen chemistry, as traced by the HCN and NH$_3$ abundances, from interstellar ices to inner protoplanetary disk surfaces. Inner disks are defined at regions within 10\,au from the central star. Also included for reference are the observed abundances in Solar System comets. The comet data points from \cite{Smyth95,Schloerb87,Meier94,Biver99,Kawakita98,Bird97,Biver12,Kawakita11,Kobayashi09,DelloRusso14}. The elliptical regions are added as a guide to indicate the approximate regions that may be populated for ices, comets and inner disks. Finally, the data points are compared to the model abundances of the inner disk surface of a T Tauri disk from \cite{Walsh15}. Note that since this figure displays ratios of observed molecular species, it is insensitive to absolute abundances and the assumed gas-to-dust ratio. The range of the outer disk NH$_3$ abundance in TW Hya from \cite{Salinas16} is also indicated.}
\label{fig:abundance_overview}
\end{figure*}

In Figure \ref{fig:abundance_overview}, we compare the observed and predicted NH$_3$ and HCN abundances in the inner disk surface for a 0.5\,$M_{\odot}$ star from \cite{Walsh15} as a function of radius. Note that this model is static, and midplane abundances may be affected by the presence of vertical mixing. It is expected that the inner disk surface has reached chemical steady state, whereas the midplane at the same radii evolves slower, resulting in significant chemical decoupling. Further, the inner disk midplane chemistry, because of the longer chemical time scale, is dependent on its initial condition.  

It is seen that the observed upper limits on the average NH$_3$ abundance are consistent with the model just outside of 0.3\,au. At radii between 0.3 and several au, the model HCN abundances are not consistent with the observations, while the NH$_3$ abundances are very low, consistent with the upper limits. The combination of the HCN and NH$_3$ data therefore supports a scenario in which we are observing relatively abundant HCN in the innermost disk within 1\,au, and that the NH$_3$ abundance is low throughout the inner disk surface, except perhaps for a very small region in the inner few tenths of an au. It also predicts that modest improvements in our sensitivity to warm NH$_3$ should lead to detections from the same gas giving rise to the observed warm HCN. 

\subsection{Potential impact on planetary composition}
This strong chemical evolution in nitrogen and other common elements has been used to trace the path of biogenic material from the interstellar medium to planetary systems. The degree of volatile depletion in planetesimals and terrestrial (exo)planets can be constrained by models \citep{Lee10, Furuya14, Schwarz14}, or by directly measuring the abundances of the main carbon, nitrogen and hydrogen carriers in protoplanetary disks \citep[e.g.,][]{Hogerheijde11, Pontoppidan14, McClure15, Min16}. For instance, the ratio of carbon to nitrogen (C/N) in comets, chondrites and the Earth was used by \cite{Bergin15} to constrain the delivery and loss of volatiles from terrestrial planets, generally finding an increasing C/N ratio as material was transported toward the disk radius forming the Earth. In this picture, the solar system exhibits evidence for increasing nitrogen volatility relative to water and carbon at smaller distances to the Sun. If a significant fraction of the inner disk nitrogen is redistributed from the less volatile NH$_3$ to HCN and N$_2$, as suggested by the observations in this work, this leads to an increase of the C/N ratio in condensible species, at least inside of a few au. This is consistent with the measured C/N ratio in chondrites and the Earth. Other mechanisms than just nitrogen volatility may also act to change the C/N ratio, such as differential loss of volatiles from parent bodies at later stages in the evolution of the planetary system \citep{Kerridge99,Marty12}. Nevertheless, the primordial branching ratio of nitrogen carriers is likely to play a key role in the volatile composition of planet-forming material.

\section{Conclusions}
We have presented new, sensitive upper limits on the amount of warm NH$_3$ gas in the inner regions of protoplanetary disks known to be rich in water. We also presented upper limits on the abundance of HCN in interstellar ices, as well as new estimates of the inner disk surface HCN abundance. We compared the relative NH$_3$ and HCN abundances in different reservoirs, tracing different evolutionary stages toward the formation of planets. We found that the observed NH$_3$ abundance relative to H$_2$O is lower than predicted by static chemical models. The observed highly depleted NH$_3$ abundances in inner disk surfaces at $\sim 1\,$au suggests that a global route for NH$_3$ destruction in the inner disk surface is efficient. However, we are not constraining the midplane abundances nor to which degree these may be affected by mixing processes. If vertical mixing is efficient, then the inner disk nitrogen reservoir is likely globally depleted in NH$_3$, leading to a higher mean volatility of the planet-forming nitrogen reservoir. This could have important consequences for the ability of the disk to deliver nitrogen to the surfaces of terrestrial planets. Future sensitive observations of warm NH$_3$ with the James Webb Space Telescope will be critical for constraining the nitrogen chemistry of planet-forming material.  

\acknowledgments
We are grateful to the referee for a constructive report that helped to clarify the paper. K.M.P. and A.B. acknowledges financial support by a NASA Origins of the Solar System grant No. OSS 11-OSS11-0120, a NASA Planetary Geology and Geophysics Program under grant NAG 5-10201. This work is based in part on observations obtained at the Gemini Observatory, which is operated by the Association of Universities for Research in Astronomy, Inc., under a cooperative agreement with the NSF on behalf of the Gemini partnership: the National Science Foundation (United States), the National Research Council (Canada), CONICYT (Chile), Ministerio de Ciencia, Tecnología e Innovación Productiva (Argentina), and Ministério da Ciência, Tecnologia e Inovação (Brazil). This work is also based in part on observations obtained at the European Southern Observatory, Paranal, Chile, within the observing program 164.I-0605. This work is also based in part on observations made with the Spitzer Space Telescope, which is operated by the Jet Propulsion Laboratory, California Institute of Technology under a contract with NASA. C.W. acknowledges support from the University of Leeds and the Science and Technology Facilities Council (grant number ST/R000549/1). This research made use of Astropy, a community-developed core Python package for Astronomy (Astropy Collaboration, 2018). This work has made use of data from the European Space Agency (ESA) mission {\it Gaia} (\url{https://www.cosmos.esa.int/gaia}), processed by the {\it Gaia} Data Processing and Analysis Consortium (DPAC, \url{https://www.cosmos.esa.int/web/gaia/dpac/consortium}). Funding for the DPAC has been provided by national institutions, in particular the institutions participating in the {\it Gaia} Multilateral Agreement. {\it Facilities:} \facility{Gemini, VLT, Spitzer}.

\bibliographystyle{apj}
\bibliography{ms}

\begin{thebibliography}{}
\expandafter\ifx\csname natexlab\endcsname\relax\def\natexlab#1{#1}\fi

\bibitem[{{Ag{\'u}ndez} {et~al.}(2018){Ag{\'u}ndez}, {Roueff}, {Le Petit}, \&
  {Le Bourlot}}]{Agundez18}
{Ag{\'u}ndez}, M., {Roueff}, E., {Le Petit}, F., \& {Le Bourlot}, J. 2018,
  \aap, 616, A19

\bibitem[{{Andrews} {et~al.}(2010){Andrews}, {Wilner}, {Hughes}, {Qi}, \&
  {Dullemond}}]{Andrews10}
{Andrews}, S.~M., {Wilner}, D.~J., {Hughes}, A.~M., {Qi}, C., \& {Dullemond},
  C.~P. 2010, \apj, 723, 1241

\bibitem[{{Ansdell} {et~al.}(2016){Ansdell}, {Williams}, {van der Marel},
  {Carpenter}, {Guidi}, {Hogerheijde}, {Mathews}, {Manara}, {Miotello},
  {Natta}, {Oliveira}, {Tazzari}, {Testi}, {van Dishoeck}, \& {van
  Terwisga}}]{Ansdell16}
{Ansdell}, M., {Williams}, J.~P., {van der Marel}, N., {et~al.} 2016, \apj,
  828, 46

\bibitem[{{Asplund} {et~al.}(2009){Asplund}, {Grevesse}, {Sauval}, \&
  {Scott}}]{asplund09}
{Asplund}, M., {Grevesse}, N., {Sauval}, A.~J., \& {Scott}, P. 2009, \araa, 47,
  481

\bibitem[{{Banzatti} {et~al.}(2014){Banzatti}, {Meyer}, {Manara},
  {Pontoppidan}, \& {Testi}}]{Banzatti14}
{Banzatti}, A., {Meyer}, M.~R., {Manara}, C.~F., {Pontoppidan}, K.~M., \&
  {Testi}, L. 2014, \apj, 780, 26

\bibitem[{{Banzatti} \& {Pontoppidan}(2015)}]{Banzatti15}
{Banzatti}, A., \& {Pontoppidan}, K.~M. 2015, \apj, 809, 167

\bibitem[{{Barentine} \& {Lacy}(2012)}]{Barentine12}
{Barentine}, J.~C., \& {Lacy}, J.~H. 2012, \apj, 757, 111

\bibitem[{{Bergin} {et~al.}(2015){Bergin}, {Blake}, {Ciesla}, {Hirschmann}, \&
  {Li}}]{Bergin15}
{Bergin}, E.~A., {Blake}, G.~A., {Ciesla}, F., {Hirschmann}, M.~M., \& {Li}, J.
  2015, Proceedings of the National Academy of Science, 112, 8965

\bibitem[{{Bird} {et~al.}(1997){Bird}, {Janardhan}, {Wilson}, {Huchtmeier},
  {Gensheimer}, \& {Lemme}}]{Bird97}
{Bird}, M.~K., {Janardhan}, P., {Wilson}, T.~L., {et~al.} 1997, Earth Moon and
  Planets, 78, 21

\bibitem[{{Bisschop} {et~al.}(2006){Bisschop}, {Fraser}, {{\"O}berg}, {van
  Dishoeck}, \& {Schlemmer}}]{Bisschop06}
{Bisschop}, S.~E., {Fraser}, H.~J., {{\"O}berg}, K.~I., {van Dishoeck}, E.~F.,
  \& {Schlemmer}, S. 2006, \aap, 449, 1297

\bibitem[{{Biver} {et~al.}(1999){Biver}, {Bockel{\'e}e-Morvan}, {Crovisier},
  {Davies}, {Matthews}, {Wink}, {Rauer}, {Colom}, {Dent}, {Despois}, {Moreno},
  {Paubert}, {Jewitt}, \& {Senay}}]{Biver99}
{Biver}, N., {Bockel{\'e}e-Morvan}, D., {Crovisier}, J., {et~al.} 1999, \aj,
  118, 1850

\bibitem[{{Biver} {et~al.}(2012){Biver}, {Crovisier}, {Bockel{\'e}e-Morvan},
  {Szutowicz}, {Lis}, {Hartogh}, {de Val-Borro}, {Moreno}, {Boissier},
  {Kidger}, {K{\"u}ppers}, {Paubert}, {Dello Russo}, {Vervack}, \&
  {Weaver}}]{Biver12}
{Biver}, N., {Crovisier}, J., {Bockel{\'e}e-Morvan}, D., {et~al.} 2012, \aap,
  539, A68

\bibitem[{{Blake} \& {Boogert}(2004)}]{Blake04}
{Blake}, G.~A., \& {Boogert}, A.~C.~A. 2004, \apjl, 606, L73

\bibitem[{{Blevins} {et~al.}(2016){Blevins}, {Pontoppidan}, {Banzatti},
  {Zhang}, {Najita}, {Carr}, {Salyk}, \& {Blake}}]{Blevins16}
{Blevins}, S.~M., {Pontoppidan}, K.~M., {Banzatti}, A., {et~al.} 2016, \apj,
  818, 22

\bibitem[{{Boogert} {et~al.}(2002){Boogert}, {Hogerheijde}, \&
  {Blake}}]{Boogert02}
{Boogert}, A.~C.~A., {Hogerheijde}, M.~R., \& {Blake}, G.~A. 2002, \apj, 568,
  761

\bibitem[{{Bottinelli} {et~al.}(2010){Bottinelli}, {Boogert}, {Bouwman},
  {Beckwith}, {van Dishoeck}, {{\"O}berg}, {Pontoppidan}, {Linnartz}, {Blake},
  {Evans}, \& {Lahuis}}]{Bottinelli10}
{Bottinelli}, S., {Boogert}, A.~C.~A., {Bouwman}, J., {et~al.} 2010, \apj, 718,
  1100

\bibitem[{{Brittain} {et~al.}(2007){Brittain}, {Simon}, {Najita}, \&
  {Rettig}}]{Brittain07}
{Brittain}, S.~D., {Simon}, T., {Najita}, J.~R., \& {Rettig}, T.~W. 2007, \apj,
  659, 685

\bibitem[{{Bruderer} {et~al.}(2015){Bruderer}, {Harsono}, \& {van
  Dishoeck}}]{Bruderer15}
{Bruderer}, S., {Harsono}, D., \& {van Dishoeck}, E.~F. 2015, \aap, 575, A94

\bibitem[{{Bruderer} {et~al.}(2014){Bruderer}, {van der Marel}, {van Dishoeck},
  \& {van Kempen}}]{Bruderer14}
{Bruderer}, S., {van der Marel}, N., {van Dishoeck}, E.~F., \& {van Kempen},
  T.~A. 2014, \aap, 562, A26

\bibitem[{{Carmona} {et~al.}(2014){Carmona}, {Pinte}, {Thi}, {Benisty},
  {M{\'e}nard}, {Grady}, {Kamp}, {Woitke}, {Olofsson}, {Roberge}, {Brittain},
  {Duch{\^e}ne}, {Meeus}, {Martin-Za{\"i}di}, {Dent}, {Le Bouquin}, \&
  {Berger}}]{Carmona14}
{Carmona}, A., {Pinte}, C., {Thi}, W.~F., {et~al.} 2014, \aap, 567, A51

\bibitem[{{Carr} \& {Najita}(2011)}]{Carr11}
{Carr}, J.~S., \& {Najita}, J.~R. 2011, \apj, 733, 102

\bibitem[{{Ciesla} {et~al.}(2015){Ciesla}, {Mulders}, {Pascucci}, \&
  {Apai}}]{Ciesla15}
{Ciesla}, F.~J., {Mulders}, G.~D., {Pascucci}, I., \& {Apai}, D. 2015, \apj,
  804, 9

\bibitem[{{Dello Russo} {et~al.}(2014){Dello Russo}, {Vervack}, {Kawakita},
  {Kobayashi}, {Weaver}, {Harris}, {Cochran}, {Biver}, {Bockel{\'e}e-Morvan},
  \& {Crovisier}}]{DelloRusso14}
{Dello Russo}, N., {Vervack}, R.~J., {Kawakita}, H., {et~al.} 2014, \icarus,
  238, 125

\bibitem[{{Dutrey} {et~al.}(1997){Dutrey}, {Guilloteau}, \&
  {Guelin}}]{Dutrey97}
{Dutrey}, A., {Guilloteau}, S., \& {Guelin}, M. 1997, \aap, 317, L55

\bibitem[{{Eistrup} {et~al.}(2016){Eistrup}, {Walsh}, \& {van
  Dishoeck}}]{Eistrup16}
{Eistrup}, C., {Walsh}, C., \& {van Dishoeck}, E.~F. 2016, \aap, 595, A83

\bibitem[{{Eistrup} {et~al.}(2018){Eistrup}, {Walsh}, \& {van
  Dishoeck}}]{Eistrup18}
---. 2018, \aap, 613, A14

\bibitem[{{Fayolle} {et~al.}(2016){Fayolle}, {Balfe}, {Loomis}, {Bergner},
  {Graninger}, {Rajappan}, \& {{\"O}berg}}]{Fayolle16}
{Fayolle}, E.~C., {Balfe}, J., {Loomis}, R., {et~al.} 2016, \apjl, 816, L28

\bibitem[{{Fletcher} {et~al.}(2014){Fletcher}, {Greathouse}, {Orton}, {Irwin},
  {Mousis}, {Sinclair}, \& {Giles}}]{Fletcher14}
{Fletcher}, L.~N., {Greathouse}, T.~K., {Orton}, G.~S., {et~al.} 2014, \icarus,
  238, 170

\bibitem[{{Furuya} \& {Aikawa}(2014)}]{Furuya14}
{Furuya}, K., \& {Aikawa}, Y. 2014, \apj, 790, 97

\bibitem[{{Gaia Collaboration} {et~al.}(2016){Gaia Collaboration}, {Prusti},
  {de Bruijne}, {Brown}, {Vallenari}, {Babusiaux}, {Bailer-Jones}, {Bastian},
  {Biermann}, {Evans}, \& et~al.}]{Gaia16}
{Gaia Collaboration}, {Prusti}, T., {de Bruijne}, J.~H.~J., {et~al.} 2016,
  \aap, 595, A1

\bibitem[{{Gaia Collaboration} {et~al.}(2018){Gaia Collaboration}, {Brown},
  {Vallenari}, {Prusti}, {de Bruijne}, {Babusiaux}, {Bailer-Jones}, {Biermann},
  {Evans}, {Eyer}, \& et~al.}]{Gaia18}
{Gaia Collaboration}, {Brown}, A.~G.~A., {Vallenari}, A., {et~al.} 2018, \aap,
  616, A1

\bibitem[{{Gerakines} {et~al.}(2004){Gerakines}, {Moore}, \&
  {Hudson}}]{Gerakines04}
{Gerakines}, P.~A., {Moore}, M.~H., \& {Hudson}, R.~L. 2004, \icarus, 170, 202

\bibitem[{{Greenwood} {et~al.}(2011){Greenwood}, {Itoh}, {Sakamoto}, {Warren},
  {Taylor}, \& {Yurimoto}}]{Greenwood11}
{Greenwood}, J.~P., {Itoh}, S., {Sakamoto}, N., {et~al.} 2011, Nature
  Geoscience, 4, 79

\bibitem[{{Halliday}(2013)}]{Halliday13}
{Halliday}, A.~N. 2013, \gca, 105, 146

\bibitem[{{Hartogh} {et~al.}(2011){Hartogh}, {Lis}, {Bockel{\'e}e-Morvan}, {de
  Val-Borro}, {Biver}, {K{\"u}ppers}, {Emprechtinger}, {Bergin}, {Crovisier},
  {Rengel}, {Moreno}, {Szutowicz}, \& {Blake}}]{Hartogh11}
{Hartogh}, P., {Lis}, D.~C., {Bockel{\'e}e-Morvan}, D., {et~al.} 2011, \nat,
  478, 218

\bibitem[{{He} {et~al.}(2016){He}, {Acharyya}, \& {Vidali}}]{He16}
{He}, J., {Acharyya}, K., \& {Vidali}, G. 2016, \apj, 823, 56

\bibitem[{{Heays} {et~al.}(2017){Heays}, {Bosman}, \& {van Dishoeck}}]{Heays17}
{Heays}, A.~N., {Bosman}, A.~D., \& {van Dishoeck}, E.~F. 2017, \aap, 602, A105

\bibitem[{{Heays} {et~al.}(2014){Heays}, {Visser}, {Gredel}, {Ubachs}, {Lewis},
  {Gibson}, \& {van Dishoeck}}]{Heays14}
{Heays}, A.~N., {Visser}, R., {Gredel}, R., {et~al.} 2014, \aap, 562, A61

\bibitem[{{Heinzeller} {et~al.}(2011){Heinzeller}, {Nomura}, {Walsh}, \&
  {Millar}}]{Heinzeller11}
{Heinzeller}, D., {Nomura}, H., {Walsh}, C., \& {Millar}, T.~J. 2011, \apj,
  731, 115

\bibitem[{{Herczeg} \& {Hillenbrand}(2014)}]{Herczeg14}
{Herczeg}, G.~J., \& {Hillenbrand}, L.~A. 2014, \apj, 786, 97

\bibitem[{{Hogerheijde} {et~al.}(2011){Hogerheijde}, {Bergin}, {Brinch},
  {Cleeves}, {Fogel}, {Blake}, {Dominik}, {Lis}, {Melnick}, {Neufeld},
  {Pani{\'c}}, {Pearson}, {Kristensen}, {Y{\i}ld{\i}z}, \& {van
  Dishoeck}}]{Hogerheijde11}
{Hogerheijde}, M.~R., {Bergin}, E.~A., {Brinch}, C., {et~al.} 2011, Science,
  334, 338

\bibitem[{{Horne} {et~al.}(2012){Horne}, {Gibb}, {Rettig}, {Brittain},
  {Tilley}, \& {Balsara}}]{Horne12}
{Horne}, D., {Gibb}, E., {Rettig}, T.~W., {et~al.} 2012, \apj, 754, 64

\bibitem[{{Houck} {et~al.}(2004){Houck}, {Roellig}, {Van Cleve}, {Forrest},
  {Herter}, {Lawrence}, {Matthews}, {Reitsema}, {Soifer}, {Watson}, {Weedman},
  {Huisjen}, {Troeltzsch}, {Barry}, {Bernard-Salas}, {Blacken}, {Brandl},
  {Charmandaris}, {Devost}, {Gull}, {Hall}, {Henderson}, {Higdon}, {Pirger},
  {Schoenwald}, {Sloan}, {Uchida}, {Appleton}, {Armus}, {Burgdorf},
  {Fajardo-Acosta}, {Grillmair}, {Ingalls}, {Morris}, \& {Teplitz}}]{Houck04}
{Houck}, J.~R., {Roellig}, T.~L., {Van Cleve}, J., {et~al.} 2004, in Society of
  Photo-Optical Instrumentation Engineers (SPIE) Conference Series, Vol. 5487,
  Optical, Infrared, and Millimeter Space Telescopes, ed. J.~C. {Mather},
  62--76

\bibitem[{{Ilgner} {et~al.}(2004){Ilgner}, {Henning}, {Markwick}, \&
  {Millar}}]{Ilgner04}
{Ilgner}, M., {Henning}, T., {Markwick}, A.~J., \& {Millar}, T.~J. 2004, \aap,
  415, 643

\bibitem[{{J{\o}rgensen} {et~al.}(2004){J{\o}rgensen}, {Sch{\"o}ier}, \& {van
  Dishoeck}}]{Jorgensen04}
{J{\o}rgensen}, J.~K., {Sch{\"o}ier}, F.~L., \& {van Dishoeck}, E.~F. 2004,
  \aap, 416, 603

\bibitem[{{Kama} {et~al.}(2016){Kama}, {Bruderer}, {van Dishoeck},
  {Hogerheijde}, {Folsom}, {Miotello}, {Fedele}, {Belloche}, {G{\"u}sten}, \&
  {Wyrowski}}]{Kama16}
{Kama}, M., {Bruderer}, S., {van Dishoeck}, E.~F., {et~al.} 2016, \aap, 592,
  A83

\bibitem[{{Kawakita} \& {Mumma}(2011)}]{Kawakita11}
{Kawakita}, H., \& {Mumma}, M.~J. 2011, \apj, 727, 91

\bibitem[{{Kawakita} \& {Watanabe}(1998)}]{Kawakita98}
{Kawakita}, H., \& {Watanabe}, J.-i. 1998, \apj, 495, 946

\bibitem[{{Kerridge}(1999)}]{Kerridge99}
{Kerridge}, J.~F. 1999, \ssr, 90, 275

\bibitem[{{Knez} {et~al.}(2009){Knez}, {Lacy}, {Evans}, {van Dishoeck}, \&
  {Richter}}]{Knez09}
{Knez}, C., {Lacy}, J.~H., {Evans}, II, N.~J., {van Dishoeck}, E.~F., \&
  {Richter}, M.~J. 2009, \apj, 696, 471

\bibitem[{{Kobayashi} \& {Kawakita}(2009)}]{Kobayashi09}
{Kobayashi}, H., \& {Kawakita}, H. 2009, \apj, 703, 121

\bibitem[{{Lacy} {et~al.}(2002){Lacy}, {Richter}, {Greathouse}, {Jaffe}, \&
  {Zhu}}]{Lacy02}
{Lacy}, J.~H., {Richter}, M.~J., {Greathouse}, T.~K., {Jaffe}, D.~T., \& {Zhu},
  Q. 2002, \pasp, 114, 153

\bibitem[{{Lahuis} \& {van Dishoeck}(2000)}]{Lahuis00}
{Lahuis}, F., \& {van Dishoeck}, E.~F. 2000, \aap, 355, 699

\bibitem[{{Lee} {et~al.}(2010){Lee}, {Bergin}, \& {Nomura}}]{Lee10}
{Lee}, J.-E., {Bergin}, E.~A., \& {Nomura}, H. 2010, \apjl, 710, L21

\bibitem[{{Li} {et~al.}(2013){Li}, {Heays}, {Visser}, {Ubachs}, {Lewis},
  {Gibson}, \& {van Dishoeck}}]{Li13}
{Li}, X., {Heays}, A.~N., {Visser}, R., {et~al.} 2013, \aap, 555, A14

\bibitem[{{Mandell} {et~al.}(2012){Mandell}, {Bast}, {van Dishoeck}, {Blake},
  {Salyk}, {Mumma}, \& {Villanueva}}]{Mandell12}
{Mandell}, A.~M., {Bast}, J., {van Dishoeck}, E.~F., {et~al.} 2012, \apj, 747,
  92

\bibitem[{{Marty}(2012)}]{Marty12}
{Marty}, B. 2012, Earth and Planetary Science Letters, 313, 56

\bibitem[{{McClure} {et~al.}(2015){McClure}, {Espaillat}, {Calvet}, {Bergin},
  {D'Alessio}, {Watson}, {Manoj}, {Sargent}, \& {Cleeves}}]{McClure15}
{McClure}, M.~K., {Espaillat}, C., {Calvet}, N., {et~al.} 2015, \apj, 799, 162

\bibitem[{{Meier} {et~al.}(1994){Meier}, {Eberhardt}, {Krankowsky}, \&
  {Hodges}}]{Meier94}
{Meier}, R., {Eberhardt}, P., {Krankowsky}, D., \& {Hodges}, R.~R. 1994, \aap,
  287, 268

\bibitem[{{Meijerink} {et~al.}(2009){Meijerink}, {Pontoppidan}, {Blake},
  {Poelman}, \& {Dullemond}}]{Meijerink09}
{Meijerink}, R., {Pontoppidan}, K.~M., {Blake}, G.~A., {Poelman}, D.~R., \&
  {Dullemond}, C.~P. 2009, \apj, 704, 1471

\bibitem[{{Min} {et~al.}(2016){Min}, {Bouwman}, {Dominik}, {Waters},
  {Pontoppidan}, {Hony}, {Mulders}, {Henning}, {van Dishoeck}, {Woitke},
  {Evans}, \& {Digit Team}}]{Min16}
{Min}, M., {Bouwman}, J., {Dominik}, C., {et~al.} 2016, \aap, 593, A11

\bibitem[{{Miotello} {et~al.}(2017){Miotello}, {van Dishoeck}, {Williams},
  {Ansdell}, {Guidi}, {Hogerheijde}, {Manara}, {Tazzari}, {Testi}, {van der
  Marel}, \& {van Terwisga}}]{Miotello17}
{Miotello}, A., {van Dishoeck}, E.~F., {Williams}, J.~P., {et~al.} 2017, \aap,
  599, A113

\bibitem[{{Morbidelli} {et~al.}(2000){Morbidelli}, {Chambers}, {Lunine},
  {Petit}, {Robert}, {Valsecchi}, \& {Cyr}}]{Morbidelli00}
{Morbidelli}, A., {Chambers}, J., {Lunine}, J.~I., {et~al.} 2000, Meteoritics
  and Planetary Science, 35, 1309

\bibitem[{{Moriarty} {et~al.}(2014){Moriarty}, {Madhusudhan}, \&
  {Fischer}}]{Moriarty14}
{Moriarty}, J., {Madhusudhan}, N., \& {Fischer}, D. 2014, \apj, 787, 81

\bibitem[{{Mumma} \& {Charnley}(2011)}]{Mumma11}
{Mumma}, M.~J., \& {Charnley}, S.~B. 2011, \araa, 49, 471

\bibitem[{{Najita} {et~al.}(2011){Najita}, {{\'A}d{\'a}mkovics}, \&
  {Glassgold}}]{Najita11}
{Najita}, J.~R., {{\'A}d{\'a}mkovics}, M., \& {Glassgold}, A.~E. 2011, \apj,
  743, 147

\bibitem[{{Najita} {et~al.}(2013){Najita}, {Carr}, {Pontoppidan}, {Salyk}, {van
  Dishoeck}, \& {Blake}}]{Najita13}
{Najita}, J.~R., {Carr}, J.~S., {Pontoppidan}, K.~M., {et~al.} 2013, \apj, 766,
  134

\bibitem[{{Noble} {et~al.}(2013){Noble}, {Theule}, {Borget}, {Danger},
  {Chomat}, {Duvernay}, {Mispelaer}, \& {Chiavassa}}]{Noble13}
{Noble}, J.~A., {Theule}, P., {Borget}, F., {et~al.} 2013, \mnras, 428, 3262

\bibitem[{{{\"O}berg} {et~al.}(2011{\natexlab{a}}){{\"O}berg}, {Boogert},
  {Pontoppidan}, {van den Broek}, {van Dishoeck}, {Bottinelli}, {Blake}, \&
  {Evans}}]{Oberg11}
{{\"O}berg}, K.~I., {Boogert}, A.~C.~A., {Pontoppidan}, K.~M., {et~al.}
  2011{\natexlab{a}}, \apj, 740, 109

\bibitem[{{{\"O}berg} {et~al.}(2010){{\"O}berg}, {Qi}, {Fogel}, {Bergin},
  {Andrews}, {Espaillat}, {van Kempen}, {Wilner}, \& {Pascucci}}]{Oberg10}
{{\"O}berg}, K.~I., {Qi}, C., {Fogel}, J.~K.~J., {et~al.} 2010, \apj, 720, 480

\bibitem[{{{\"O}berg} {et~al.}(2011{\natexlab{b}}){{\"O}berg}, {Qi}, {Fogel},
  {Bergin}, {Andrews}, {Espaillat}, {Wilner}, {Pascucci}, \&
  {Kastner}}]{Oberg11b}
---. 2011{\natexlab{b}}, \apj, 734, 98

\bibitem[{{Pontoppidan} {et~al.}(2011){Pontoppidan}, {Blake}, \&
  {Smette}}]{Pontoppidan11}
{Pontoppidan}, K.~M., {Blake}, G.~A., \& {Smette}, A. 2011, \apj, 733, 84

\bibitem[{{Pontoppidan} {et~al.}(2008){Pontoppidan}, {Blake}, {van Dishoeck},
  {Smette}, {Ireland}, \& {Brown}}]{Pontoppidan08}
{Pontoppidan}, K.~M., {Blake}, G.~A., {van Dishoeck}, E.~F., {et~al.} 2008,
  \apj, 684, 1323

\bibitem[{{Pontoppidan} \& {Blevins}(2014)}]{Pontoppidan14b}
{Pontoppidan}, K.~M., \& {Blevins}, S.~M. 2014, Faraday Discussions, 169, 49

\bibitem[{{Pontoppidan} {et~al.}(2009){Pontoppidan}, {Meijerink}, {Dullemond},
  \& {Blake}}]{Pontoppidan09}
{Pontoppidan}, K.~M., {Meijerink}, R., {Dullemond}, C.~P., \& {Blake}, G.~A.
  2009, \apj, 704, 1482

\bibitem[{{Pontoppidan} {et~al.}(2014){Pontoppidan}, {Salyk}, {Bergin},
  {Brittain}, {Marty}, {Mousis}, \& {{\"O}berg}}]{Pontoppidan14}
{Pontoppidan}, K.~M., {Salyk}, C., {Bergin}, E.~A., {et~al.} 2014, Protostars
  and Planets VI, 363

\bibitem[{{Pontoppidan} {et~al.}(2010{\natexlab{a}}){Pontoppidan}, {Salyk},
  {Blake}, \& {K{\"a}ufl}}]{Pontoppidan10b}
{Pontoppidan}, K.~M., {Salyk}, C., {Blake}, G.~A., \& {K{\"a}ufl}, H.~U.
  2010{\natexlab{a}}, \apjl, 722, L173

\bibitem[{{Pontoppidan} {et~al.}(2010{\natexlab{b}}){Pontoppidan}, {Salyk},
  {Blake}, {Meijerink}, {Carr}, \& {Najita}}]{Pontoppidan10}
{Pontoppidan}, K.~M., {Salyk}, C., {Blake}, G.~A., {et~al.} 2010{\natexlab{b}},
  \apj, 720, 887

\bibitem[{{Pontoppidan} {et~al.}(2003){Pontoppidan}, {Fraser}, {Dartois},
  {Thi}, {van Dishoeck}, {Boogert}, {d'Hendecourt}, {Tielens}, \&
  {Bisschop}}]{Pontoppidan03}
{Pontoppidan}, K.~M., {Fraser}, H.~J., {Dartois}, E., {et~al.} 2003, \aap, 408,
  981

\bibitem[{{Prato} {et~al.}(2003){Prato}, {Greene}, \& {Simon}}]{Prato03}
{Prato}, L., {Greene}, T.~P., \& {Simon}, M. 2003, \apj, 584, 853

\bibitem[{{Qi} {et~al.}(2004){Qi}, {Ho}, {Wilner}, {Takakuwa}, {Hirano},
  {Ohashi}, {Bourke}, {Zhang}, {Blake}, {Hogerheijde}, {Saito}, {Choi}, \&
  {Yang}}]{Qi04}
{Qi}, C., {Ho}, P.~T.~P., {Wilner}, D.~J., {et~al.} 2004, \apjl, 616, L11

\bibitem[{{Rice} {et~al.}(2018){Rice}, {Bergin}, {J{\o}rgensen}, \&
  {Wampfler}}]{Rice18}
{Rice}, T.~S., {Bergin}, E.~A., {J{\o}rgensen}, J.~K., \& {Wampfler}, S.~F.
  2018, \apj, 866, 156

\bibitem[{{Salinas} {et~al.}(2016){Salinas}, {Hogerheijde}, {Bergin},
  {Cleeves}, {Brinch}, {Blake}, {Lis}, {Melnick}, {Pani{\'c}}, {Pearson},
  {Kristensen}, {Y{\i}ld{\i}z}, \& {van Dishoeck}}]{Salinas16}
{Salinas}, V.~N., {Hogerheijde}, M.~R., {Bergin}, E.~A., {et~al.} 2016, \aap,
  591, A122

\bibitem[{{Salyk} {et~al.}(2013){Salyk}, {Herczeg}, {Brown}, {Blake},
  {Pontoppidan}, \& {van Dishoeck}}]{Salyk13}
{Salyk}, C., {Herczeg}, G.~J., {Brown}, J.~M., {et~al.} 2013, \apj, 769, 21

\bibitem[{{Salyk} {et~al.}(2014){Salyk}, {Pontoppidan}, {Corder}, {Mu{\~n}oz},
  {Zhang}, \& {Blake}}]{Salyk14}
{Salyk}, C., {Pontoppidan}, K., {Corder}, S., {et~al.} 2014, \apj, 792, 68

\bibitem[{{Salyk} {et~al.}(2011){Salyk}, {Pontoppidan}, {Blake}, {Najita}, \&
  {Carr}}]{Salyk11}
{Salyk}, C., {Pontoppidan}, K.~M., {Blake}, G.~A., {Najita}, J.~R., \& {Carr},
  J.~S. 2011, \apj, 731, 130

\bibitem[{{Schloerb} {et~al.}(1987){Schloerb}, {Kinzel}, {Swade}, \&
  {Irvine}}]{Schloerb87}
{Schloerb}, F.~P., {Kinzel}, W.~M., {Swade}, D.~A., \& {Irvine}, W.~M. 1987,
  \aap, 187, 475

\bibitem[{{Schwarz} \& {Bergin}(2014)}]{Schwarz14}
{Schwarz}, K.~R., \& {Bergin}, E.~A. 2014, \apj, 797, 113

\bibitem[{{Semenov} \& {Wiebe}(2011)}]{Semenov10}
{Semenov}, D., \& {Wiebe}, D. 2011, \apjs, 196, 25

\bibitem[{{Smyth} {et~al.}(1995){Smyth}, {Combi}, {Roesler}, \&
  {Scherb}}]{Smyth95}
{Smyth}, W.~H., {Combi}, M.~R., {Roesler}, F.~L., \& {Scherb}, F. 1995, \apj,
  440, 349

\bibitem[{{Tart{\`e}se} \& {Anand}(2013)}]{Tartese13}
{Tart{\`e}se}, R., \& {Anand}, M. 2013, Earth and Planetary Science Letters,
  361, 480

\bibitem[{{Thi} {et~al.}(2013){Thi}, {Kamp}, {Woitke}, {van der Plas},
  {Bertelsen}, \& {Wiesenfeld}}]{Thi13}
{Thi}, W.~F., {Kamp}, I., {Woitke}, P., {et~al.} 2013, \aap, 551, A49

\bibitem[{{van Dishoeck} {et~al.}(2003){van Dishoeck}, {Dartois},
  {Pontoppidan}, {Thi}, {D'Hendecourt}, {Boogert}, {Fraser}, {Schutte}, \&
  {Tielens}}]{Dishoeck03}
{van Dishoeck}, E.~F., {Dartois}, E., {Pontoppidan}, K.~M., {et~al.} 2003, The
  Messenger, 113, 49

\bibitem[{{Walsh} {et~al.}(2015){Walsh}, {Nomura}, \& {van Dishoeck}}]{Walsh15}
{Walsh}, C., {Nomura}, H., \& {van Dishoeck}, E. 2015, \aap, 582, A88

\bibitem[{{Zhang} {et~al.}(2013){Zhang}, {Pontoppidan}, {Salyk}, \&
  {Blake}}]{Zhang13}
{Zhang}, K., {Pontoppidan}, K.~M., {Salyk}, C., \& {Blake}, G.~A. 2013, \apj,
  766, 82

\bibitem[{{Zhang} {et~al.}(2018){Zhang}, {Zhu}, {Huang}, {Guzm{\'a}n},
  {Andrews}, {Birnstiel}, {Dullemond}, {Carpenter}, {Isella}, {P{\'e}rez},
  {Benisty}, {Wilner}, {Baruteau}, {Bai}, \& {Ricci}}]{Zhang18}
{Zhang}, S., {Zhu}, Z., {Huang}, J., {et~al.} 2018, \apjl, 869, L47

\end{thebibliography}

\end{document}